\documentclass[aps,pra,superscriptaddress,amsmath,showpacs,tightenlines]{revtex4-2}

\usepackage{amsmath}
\usepackage{amssymb}
\usepackage{graphicx}
\usepackage{bm}
\usepackage{color}
\usepackage{epsfig}
\usepackage{amsfonts}

\begin{document}

\title{Frequency adjustable Resonator as a Tunable Coupler for Xmon Qubits}

\author{Hui Wang}
\affiliation{Inspur Academy of science and technology, Jinan, China}
\affiliation{Inspur artificial intelligence research institute, Jinan, China}

 \author{Yan-Jun Zhao}
\email{Correspondence:zhao_yanjun@bjut.edu.cn}
\affiliation{Key Laboratory of Opto-electronic Technology, Ministry of Education, Beijing University of
Technology, Beijing, China}

\author{Rui Wang}
\affiliation{ Department of Physics, Tokyo University of Science, 1–3 Kagurazaka, Shinjuku, Tokyo 162–0825, Japan. }
\affiliation{ RIKEN Center for Quantum Computing (RQC), Wako, Saitama 351-0198, Japan}

\author{Xun-Wei Xu}
\affiliation{Key Laboratory of Low-Dimensional Quantum Structures and Quantum Control of Ministry of Education,
 Key Laboratory for Matter Microstructure and Function of Hunan Province,
  Department of Physics and Synergetic Innovation Center for Quantum Effects and Applications,
   Hunan Normal University, Changsha 410081, China}

\author{Qiang Liu}
\affiliation{Inspur Academy of science and technology, Jinan, China}
\affiliation{Inspur artificial intelligence research institute quantum computer}

\author{Jianhua Wang}
\affiliation{Inspur Academy of science and technology, Jinan, China}

\author{Changxin Jin}
\affiliation{Inspur Academy of science and technology, Jinan, China}

\begin{abstract}
We propose a scheme of tunable coupler based on a frequency adjustable  resonator for scalable quantum integrated circuits. One side of the T-shape quarter-wave resonator is short to the ground through a DC SQUID, which is used to tune the frequency of the resonator coupler. In the other side with two open ends, either end  capacitively couples to a different Xmon qubit, which  determines the effective  coupling between the two qubits.
The fundamental mode of the tunable resonator dominates the interaction with the qubits and  functions as a tunable coupler to switch off/on the qubit-qubit interaction. The required external magnetic flux for the tunable coupler's  switching off frequency is affected by the asymmetry degree of the DC SQUID, which can help to choose a better noise environment (of the qubits) induced by the resonator coupler. The  T-shape tunable resonator can help to reduce the  direct interaction between two qubits, which should be an important way to suppress the ZZ crosstalk and realize high-fidelity quantum gates. The  DC SQUID is several millimeters away from the Xmon qubits, which in principle creates less flux noises compared with the transmon-type coupler.
\end{abstract}
\date{\today}

\maketitle

\section{Introduction}

High-quality two-qubit quantum gates are the key step for future large-scale  superconducting quantum processor
and quantum error correction. Dynamically tuning  the qubit-qubit coupling greatly enhances the fidelity of two qubit quantum gate, and the schemes include the tunable inductance \cite{Chen}, transmon coupler\cite{Yan,Sun},
 rf-SQUID\cite{Allman}, resonator coupler\cite{Ming,IBM,Kandala}, floating coupler\cite{Sete}, and so on.
  In the past few years, the high-precision control of  multiple qubits on superconducting quantum chips
   have been experimentally realized\cite{Ming,IBM,Martinis,Wu,Friis,Mooney,Zhang}.
  But the residual  ZZ crosstalk and state leakage  still  limit for further improvement
on control precision for superconducting quantum processor.
 Many experimental schemes have been used to suppress the ZZ crosstalk and
   realize high-fidelity two-qubit quantum gates\cite{Chen,Kandala,Sete,Zhao,Tan,Sung,Mundada,Barends}.

The superconducting resonator is easy to fabricate and measure, which has been used as the coupler in
 the superconducting quantum processor. For example,  IBM have developed the  'Eagle' quantum processor
 with to date the largest number of 127 transmon qubits\cite{Ming,IBM}.
 But in most experiments, the frequencies of resonator-type couplers are fixed,
  and  the qubit-qubit interaction can not be totally switched off.
  By embedding a DC SQUID inside the superconducting resonator,
 some experiments have tuned resonator's frequency by more than 500 Megahertz,
and the tuning speed  can be faster than the lifetime of photons\cite{Cleland,Sandberg,Yamaji,Wustmann1,Delsing,IDA}.
  Recently the exponentially large on-off ratio has been realized in experiment with a nonlinear
mode of resonator  as the tunable superconducting coupler \cite{Leroux}.

In this paper, we propose a theoretical scheme of using a T-shape quarter-wave resonator
 to dynamically tune the coupling between two Xmon qubits.
The open ends of T-shape resonator capacitively couple to two qubits,
 while the other end is a DC SQUID which dominates the inductive energy of the resonator coupler.
 Considering the tunable frequency  with the external magnetic flux and  largely non-equidistant basic modes ,
     the fundamental mode of quarter-wave resonator could function as  a tunable coupler to switch off/on the qubit-qubit coupling.
    For the asymmetric DC SQUID, the required external magnetic flux for the fundamental mode to reach the switch off frequency can be farther from the half-integer quantum flux, and the degree of asymmetry could help to choose a  better noise environment for the Xmon qubits.
With the
  transversely broaden part of T-type resonator at the open to weaken the direct qubit-qubit coupling,  the  ZZ crosstalk
 can be suppressed to several kilohertz which  guarantees the realizability of the high fidelity quantum gates.
 And the DC SQUID is almost 5 mm away and needs a small current to control the magnetic flux,
  so the resonator coupler should create less flux noises to  Xmon qubits.
  Compared with the transmon type coupler, the resonator-based coupler is easily to  fabricate and measure, and has the potential  advantages of saving the  dilution refrigerator lines.

The paper is organized as follows: In Sec.~II, we discuss the
theoretical model of a three-body system, In Sec.~III,
we analyze the cavity modes' frequencies and anharmonicities.
In Sec.~IV, we study the effective qubit-qubit coupling,  ZZ crosstalk,
 and state leakage to the resonator coupler.  We finally summarize the results in Sec.~V.

\section{The Physical Model}

\begin{figure}
\centering\includegraphics[bb=0 60 410 445, width=8.0cm, clip]{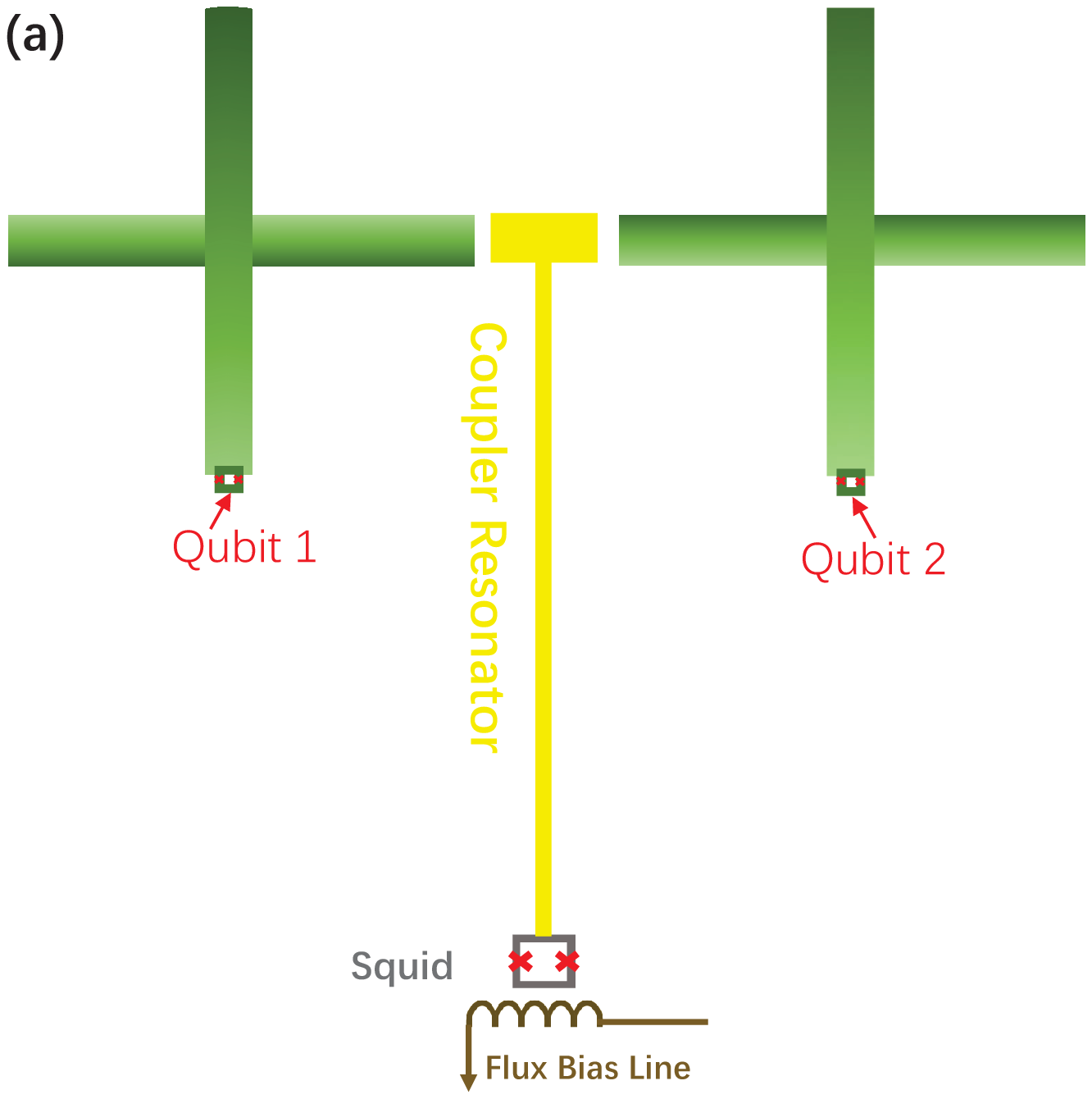}\\
\centering\includegraphics[bb=15 270 150 715,    width=2.55 cm, angle=90, clip]{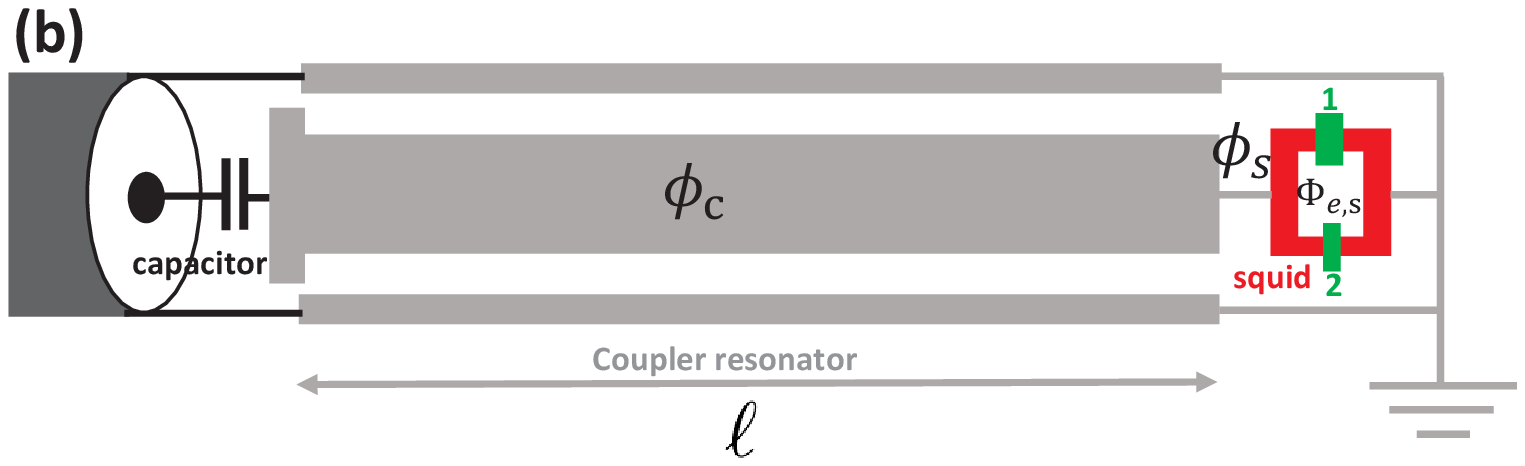}
\caption{(Color online) T-type quarter-wave resonator functioning as the tunable  coupler for two Xmon qubits.
(a) Diagram sketch of the three-body system. One side of the resonator with the two ends capacitively couples to two different qubits, and the other side is short to the ground through a DC SQUID. The transverse width of T-shape resonator is much smaller than the longitudinal length ($\ell$).
(b) Detailed sketch of the T-shape quarter-wave resonator\cite{IDA}. Here $\phi_c$ is the superconducting phase of resonator, while $\phi_s$ is the boundary value of the cavity field at the SQUID, and
 the external magnetic flux $\Phi_{e,s}$ could tune the frequency of the resonator coupler.
}
\label{fig1}
\end{figure}

As shown in Fig.~\ref{fig1}(a), we study a three-body system consisting of two qubits capacitively coupling to
 a common T-shape quarter-wave resonator, the DC SQUID is far away and does not participate in the interaction with qubits.
 Then the Hamiltonian of three-body system can be written as
\begin{eqnarray}\label{eq:1}
H&=&\sum^2_{j=1}\frac{\hbar}{2}\omega_{j}\sigma^{z}_{j}+\sum_{n}\hbar\omega^{(n)}_c(\Phi_{e,s}) a^{\dagger}_n a_n\\
& +&\sum_{j,n}\frac{\hbar}{2} g^{(n)}_{jc}[\sigma^{+}_{j}a_n+\sigma^{-}_{j}a^{\dagger}_n]+\hbar g_{12}[\sigma^{-}_{1}\sigma^{+}_{2}+\sigma^{-}_{2}\sigma^{+}_{1}],\nonumber
\end{eqnarray}
here $\omega_{j}$ are the transition frequencies of qubits, where the subsrib  $j=1,2$.
 The symbols $a^{\dagger}_n$ and $a_n$ are the creation and annihilation operators of the $n$-th resonator mode's photons,
  while $\sigma^{\pm}_{j}$ and $\sigma^{z}_{j}$
 are the ladder operators and Pauli Z operators of the two qubits, respectively.
And  $\omega^{(n)}_c$ ($n$ are integers) is the $n$-th mode of the T-shape quarter-wave resonator. As will be discussed in the following Sections, the inductive energy  of DC SQUID is much larger than that of transmission line resonator,
 and the frequency of the quarter-wave resonator can be continually tuned by the external magnetic flux $\Phi_{e,s}$.
The quantity $g_{12}$ is the direct coupling strength between two qubits,
and  $g^{(n)}_{jc}$ describe  coupling strengths of  qubits with $n$-th  resonator mode.
The qubit-resonator coupling strength $g^{(n)}_{jc}$ also depends on  external magnetic flux $\Phi_{e,s}$
as will be discussed in the following Sections.

The switching off frequency for qubit-qubit coupling usually locates in the qubit-coupler large detuning regime, that is $g^{(n)}_{jc}/|\Delta^{(n)}_j|\ll 1$,
where $\Delta^{(n)}_{j}=\omega_{j}-\omega^{(n)}_c$.
 Defining $S=\sum_{j,n}(g^{(n)}_{jc}/\Delta^{(n)}_j)[a^{\dagger}_n\sigma^{(j)}_{-}-a_n\sigma^{(j)}_{+}]$, under the unitary transformation $H^{\prime}=\exp(-S)H\exp(S)$, then we get the effective Hamiltonian in the large detuning regime as
 \begin{eqnarray}\label{eq:2}
H^{\prime}&\approx& \sum_{n}\hbar\left[\omega^{(n)}_c+\sum^{2}_{j=1}\frac{\left[g^{(n)}_{jc}\right]^2}{\Delta^{(n)}_{j}}\sigma^z_j\right]a^{\dagger}_n a_n\nonumber\\
& +&\frac{\hbar}{2}\sum_{j}\left[\omega_{j}+\sum^{2}_{n=1}\frac{\left[g^{(n)}_{jc}\right]^2}{\Delta^{(n)}_{j}}\right]\sigma^{z}_{j}\nonumber\\
& +&\hbar\left[g_{12}+\sum_{n}\frac{g^{(n)}_{1c}g^{(n)}_{2c}}{\Delta^{(n)}_{e}}\right]\left[\sigma^{-}_{1}\sigma^{+}_{2}+\sigma^{-}_{2}\sigma^{+}_{1}\right].
\end{eqnarray}
Here, $2/\Delta^{(n)}_{e}=1/\Delta^{(n)}_1+1/\Delta^{(n)}_2$, and the condition $g_{12}\ll g^{(n)}_{jc}\ll |\Delta^{(n)}_j|$ has been used above.    The representation $ (g^{(n)}_{1c}g^{(n)}_{2c})/\Delta^{(n)}_{e}$ describes the amplitude of indirect qubit-qubit coupling induced by the $n$-th resonator mode, and the qubit-qubit coupling is switched off
when $\sum_{n}g^{(n)}_{1c}g^{(n)}_{2c}/\Delta^{(n)}_{e}=-g_{12}$, here $n$ is an integer.



   As will be discussed in the following Sections, the spectrum anharmonicities of T-shape quarter-wave resonator are very large and can reach Gigahertz,  so the  modes hybridization induced by qubits could be neglected. So the fundamental mode of the tunable resonator will dominate the interactions with the qubits and functions as a tunable coupler for the superconducting qubits.
  If the  frequency $\omega^{(1)}_c(\Phi_{e,s})$ is tuned to a certain frequency to satisfy $g^{(1)}_{1c}g^{(1)}_{2c}/\Delta^{(1)}_{e}=-g_{12}$, then the qubit-qubit coupling can be totally switched off for an isolated single-qubit quantum gate. Moving away from this frequency, the qubit-qubit coupling will be switched on to allow the multi-qubit quantum gates.

\section{T-shape Quarter-wave Resonator}

 Figure \ref{fig1}(a) describes a three-body system consisting of two Xmon qubits and a T-shape quarter-wave resonator, the fundamental mode of the resonator functions as a tunable coupler for the qubit-qubit coupling.
 The detailed sketch of tunable resonator is  shown in Fig.~\ref{fig1}(b), it is a deformed quarter-wave resonator with a transversely broadened part at the open end (left side), and  the other end (right side) is short to the ground through a DC SQUID. The longitudinal length of transmission line resonator is $\ell$ (about 4.87mm ), which is much larger than the transverse width (usually below 100$\mu$m), so the effects of the transversely broadened part  on resonator coupler's basic modes can be neglected.

The capacitance and inductance  per unit length are $C_0$ and $L_0$, respectively.
The  DC SQUID  consists of two junctions with critical currents
$I_{cs_{1,2}}$, respectively, and $C_{s_{1,2}}$ are their capacitances.
The quantities $\phi_{s_{1,2}}$ are  the superconducting phases across the junctions 1 and 2 of the DC SQUID, respectively,
and they satisfy the relation $\phi_{s_1}-\phi_{s_2}=2\pi n+2\pi\Phi_{e,s}/\Phi_0$ ($n$ is an integer).
 Then the Lagrangian of the T-shape quarter-wave resonator is
\begin{eqnarray}\label{eq:3}
L_{qwr}&=&\left(\frac{\Phi_0}{2\pi}\right)^2\int^{\ell}_{0}\left[\frac{C_{0}}{2} \dot{\phi }^2_c(x,t)-\frac{\left[\partial_x \phi_c(x,t)\right]^2}{2L_0 }\right]dx\nonumber\\
& &+\frac{\hbar^2}{8E_{C_{s_1}}}\dot{\phi}^2_{s_1}+E_{J{s_1}} \cos(\phi_{s_1})\nonumber\\
& &+\frac{\hbar^2}{8E_{C_{s_2}}}\dot{\phi}^2_{s_2}+ E_{J{s_2}} \cos(\phi_{s_2}).
\end{eqnarray}
 Here $E_{C{s_{1,2}}}=e^2/(2C_{s_{1,2}})$
 and  $E_{J{s_{1,2}}}=-\Phi_0 I_{cs_{1,2}}/(2\pi)$  describe the charging energies and Josephson  energies of SQUID's two junctions, respectively.
Here $\phi_c(x,t)$ is the superconducting phase operator of the resonator, and $\Phi_0=h/2e$ is the flux quantum.
 If we define $E_{Jt}=E_{Js_1}+E_{Js_2}$, then the effective Josephson energy of the DC SQUID can be obtained as $E_{Js}=E_{Jt}\sqrt{\cos(\pi\Phi_{e,s}/\Phi_0)^2+d^2\sin^2(\pi\Phi_{e,s}/\Phi_0)}$, where  $d=(E_{Js_1}-E_{Js_2})/(E_{Js_1}+E_{Js_2})$  describes the junction asymmetry\cite{Schuster}.
 The DC SQUID can be regarded as an inductor with the tunable inductance,
 \begin{eqnarray}\label{eq:4}
  L_{sq}=\frac{\Phi_{0}/(2\pi I_{cs})}{\sqrt{\cos^2\left(\frac{\pi\Phi_{e,s}}{\Phi_0}\right)+d^2\sin^2\left(\frac{\pi\Phi_{e,s}}{\Phi_0}\right)} \cos(\phi_s-\phi_0)},
 \end{eqnarray}
here $I_{cs}=I_{{cs}_1}+I_{{cs}_2}$ is the  sum of the critical currents of the two junctions,
and the phase $\phi_0$ can be solved from the equation: $\tan \phi_0 =d\tan\left(\pi\Phi_{e,s}/\Phi_0\right)$.
The boundary value of cavity field at the SQUID is defined as $\phi_s=(\phi_{s_1}+\phi_{s_2})/2$
 which can be considered as the resonator's superconducting phase  at the SQUID.
 For typical dimensions of devices, the Josephson energy of the DC SQUID is much larger
  than that of transmission line resonator and dominates the inductive energy of the resonator coupler,
  so the frequency of the resonator coupler can be tuned by external magnetic flux $\Phi_{e,s}$ passing through the SQUID loop.
 Some experiments have realized fast tuning of resonator frequency
   exceeding  500 Megahertz and even  one Gigahertz\cite{Cleland,Sandberg,Yamaji}.

\begin{figure}
\centering\includegraphics[bb=0 0 465 420, width=4.10 cm, clip]{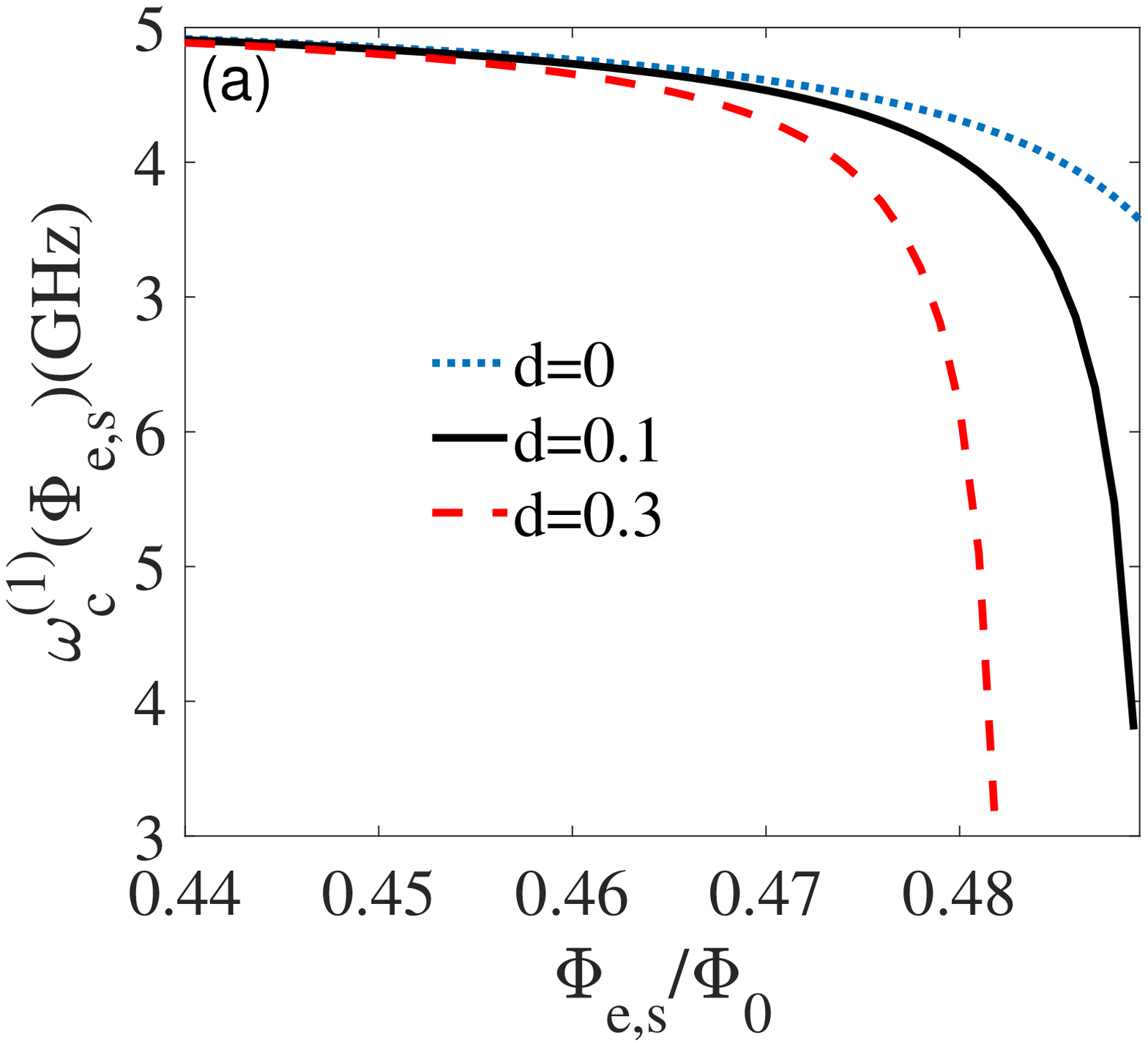}
\centering\includegraphics[bb=0 0 480 425, width=4.12 cm, clip]{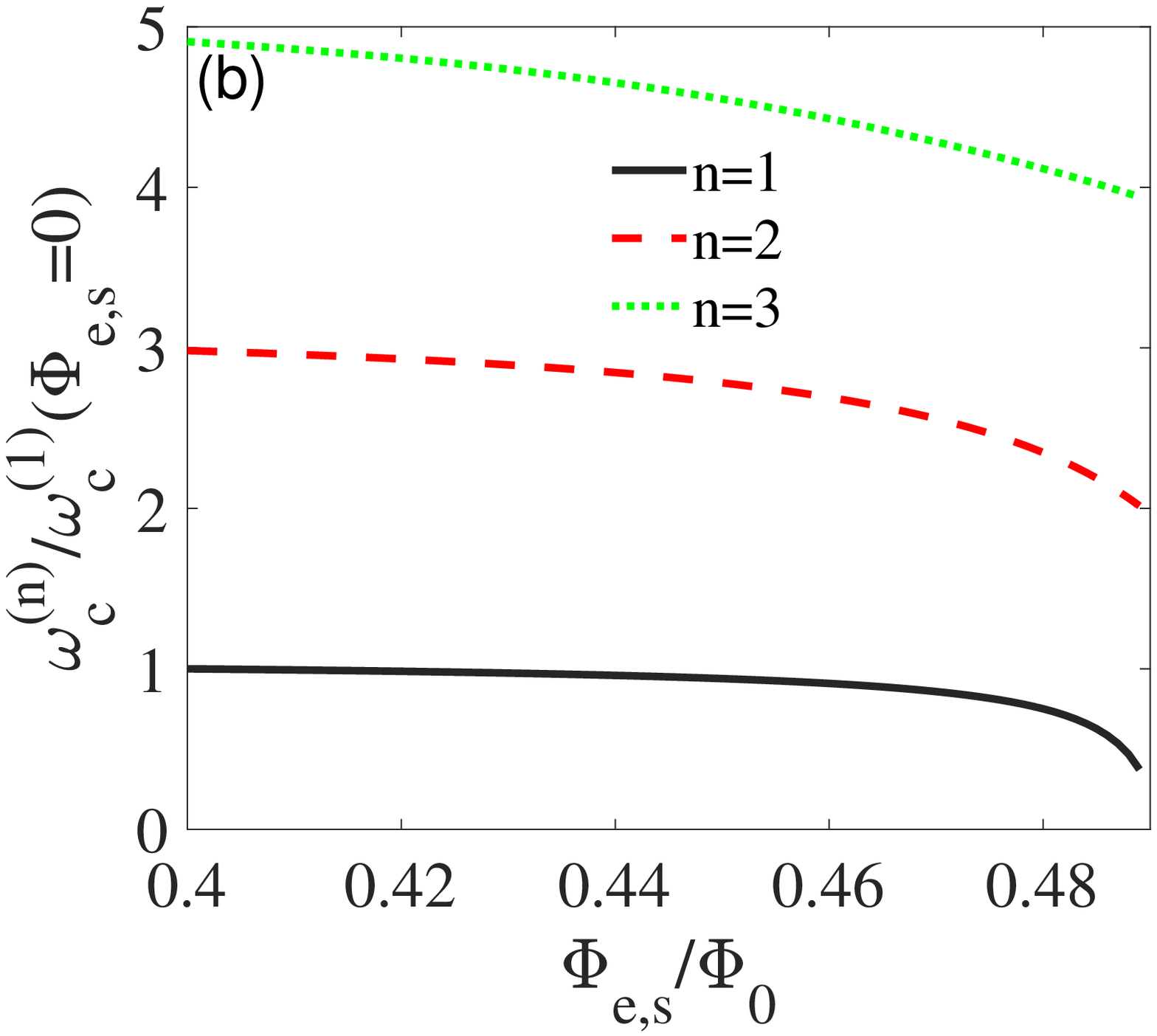}\\
\centering\includegraphics[bb=0 0 475 400, width=4.25 cm, clip]{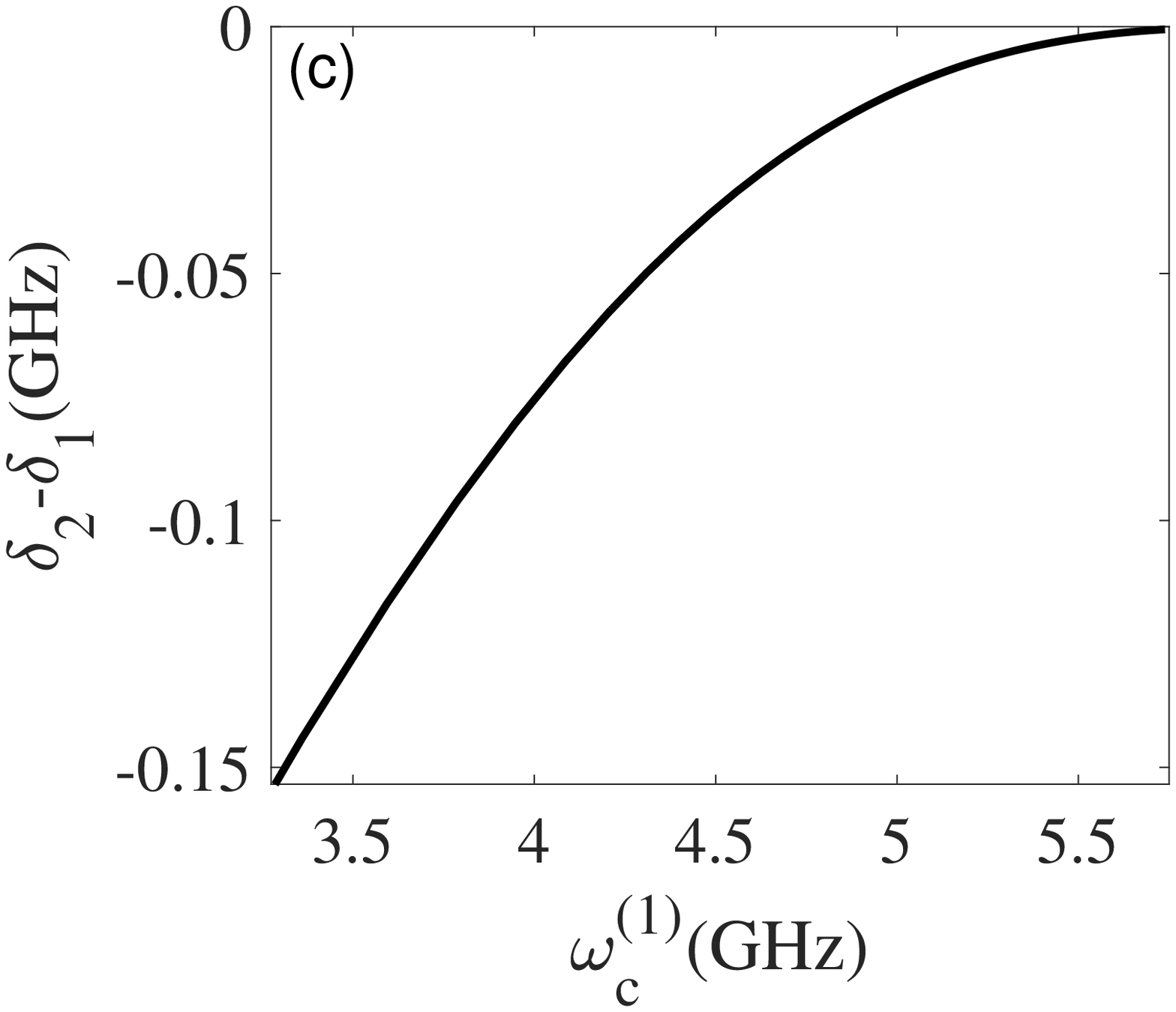}
\centering\includegraphics[bb=5 0  500 420, width=4.30 cm, clip]{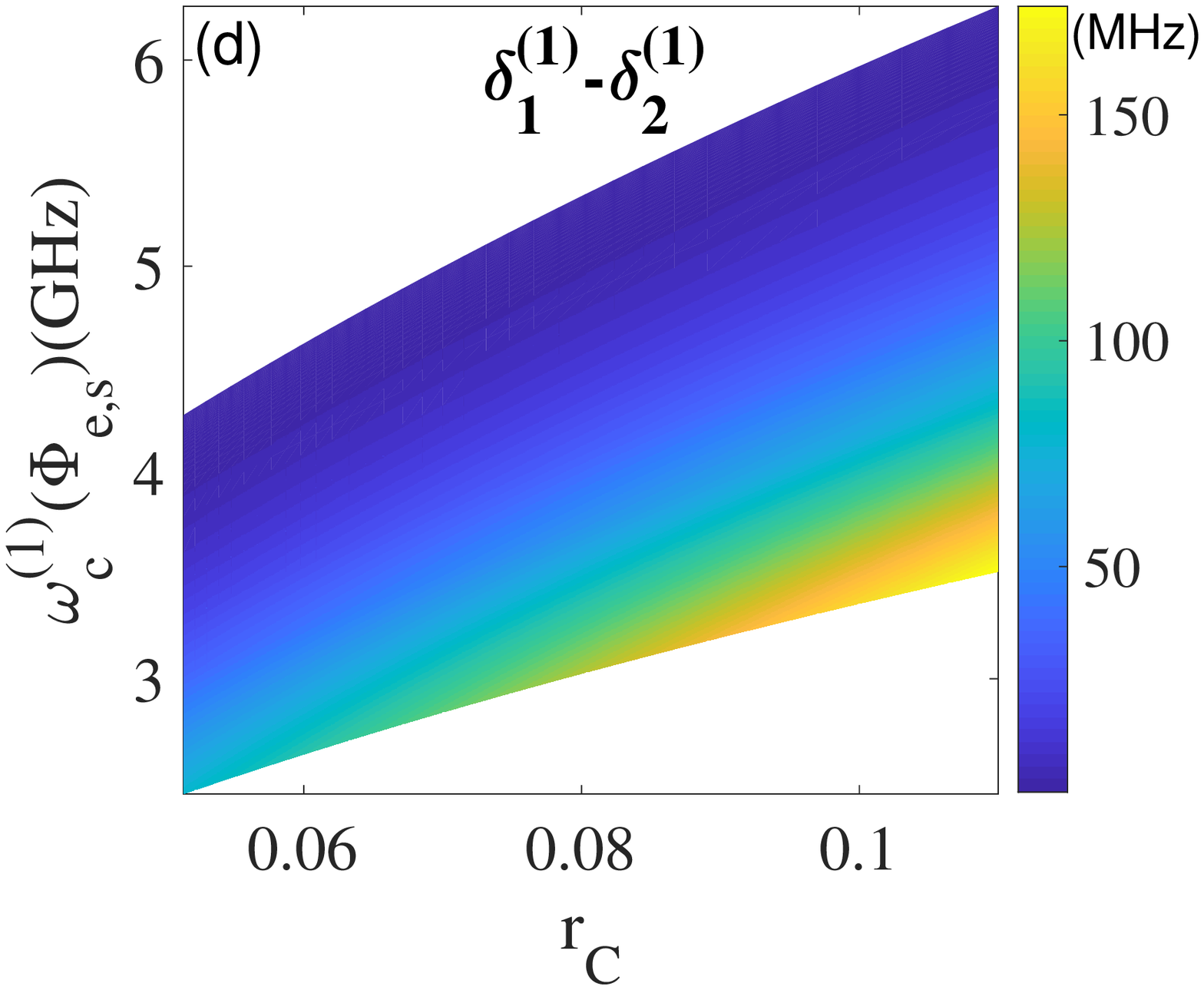}
\caption{(Color online) Frequencies and  anharmonicities for the tunable resonator's basic modes.
(a) Fundamental mode's frequencies $\omega^{(1)}_c$ changing with the external magnetic flux $\Phi_{e,s}$ under different asymmetries $d$ of the SQUID; (b) three lowest basic modes' frequencies $\omega^{(n)}_c$ ($n=1,2,3$) changing with the  magnetic flux $\Phi_{e,s}$;
(c) anharmonicity ($\delta^{(1)}_2-\delta^{(1)}_1$) of fundamental mode's two lowest energy levels changing with $\omega^{(1)}_c$;
(d) anharmonicity ($\delta^{(1)}_2-\delta^{(1)}_1$) changing
with the frequency $\omega^{(1)}_c$ and the capacitance $(C_0\ell)$, with $r_C=C_s/C_0 \ell$.
The longitudinal length of  resonator coupler $\ell=4.87$ mm, the capacitance  per unit length  $C_0=0.16$ nF/m [except in (d)]
  and inductance  per unit length $L_0=0.44$ $\mu$H/m.
 The other parameters are: $r_L=0.02$, $r_C=0.1$, $d=0.1$ [except in (a)], the SQUID's inductance $C_s=75$ fF, and the fundamental mode's frequency of the bare resonator  is $\omega^{(b)}_c(\Phi_{e,s}=0)=6$ GHz [except in (d)].
 }
\label{fig2}
\end{figure}

The resonator's superconducting phase field $\phi_c$ follows
 the wave equation $\ddot{\phi}_c-v^2\phi^{\prime\prime}_c=0$,
 which can be expanded with the resonator's eigen-modes,
 then yielding $\phi_c(x,t)=(2e/\hbar)\sqrt{(2/C_c \ell)}\sum_{n}q_n(t)\cos(k_n x)$ ($n$ are integers).
Here $\omega^{(n)}_c$  is the frequency of $\textit{n}$-th basic mode,
  $q_n(t)$ are time-dependent coefficients and $k_n=\omega^{(n)}_c/v$,
  and $v=1/\sqrt{L_0 C_0}$  is the phase velocity.
    At the open end  $ \partial_x \phi_c(x,t)|_{x=0}=0$,
 while the boundary value of cavity field at the SQUID satisfies $\phi_s(x,t)=\phi_c(x,t)|_{x=\ell}$,
    the motion equation of $\phi_s$ can be written as
    $\hbar^2 \ddot{\phi}_s/E_{C_s}+2E_{Js} \cos(\pi\Phi_{e,s}/\Phi_0)\sin(\phi_s)+\ell E_{L,cav} \phi^{\prime}_s=0$,
  where  $\phi^{\prime}_s=\partial_x \phi_s(x,t)$, and $E_{L,cav}=(\hbar/2e)^2 /(L_0 \ell)$ is the inductive
  energy of transmission line resonator\cite{Delsing,Wustmann1,IDA}.
  In this paper we focus on the regime  $|\phi_s|\ll1$,
   and the $\phi_s$ decouples from other variables of quarter-wave resonator.
Under the static bias current, the superconducting phase field can be formally written as $\phi_c(x,t)\varpropto \cos(k_n x)\exp(\pm i\omega_n t)$.
 Substituting it into the motion equation of $\phi_s$,
 gives the dispersion relation for the $n$-th mode, i.e.,
\begin{eqnarray}\label{eq:5}
0&=&k_n \ell \tan(k_n \ell)+ r_C (k_n \ell)^2-\frac{\cos(\phi_s-\phi_0)}{r_L}\nonumber\\
& &\times\sqrt{\cos^2\left(\frac{\pi\Phi_{e,s}}{\Phi_0}\right)+d^2\sin^2\left(\frac{\pi\Phi_{e,s}}{\Phi_0}\right)}.
\end{eqnarray}
Here,  $r_C=C_s/C_0 \ell$ is the capacitance ratio between the DC SQUID and the transmission line resonator, and  $r_L=E_{l,cav}/E_{Js}$ is the  ratio of inductive energies between the transmission line resonator and the DC SQUID.     In this paper we focus on the  regime $E_{l,cav}/[2E^{(m)}_{Js}\cos(\pi\Phi_{e,s}/\Phi_{0})]\ll 1$ and $|\phi_{s}|\ll 1$,
  which guarantee the  SQUID's variables decoupling from the resonator modes, here $E^{(m)}_{Js}=-\Phi_0 I_{cs}/(2\pi)$ maximal Josephson energy of DC SQUID.   The numerical simulation results from Eq.~(\ref{eq:5}) can  be seen  in Fig.~\ref{fig2}. The blue-dotted curve  in Fig.~\ref{fig2}(a) describes the variations of the fundamental mode's frequency of symmetric DC SQUID ($d=0$), and the frequency of fundamental mode can be effectively tuned by the external magnetic flux. For the asymmetric DC SQUID, the fundamental mode's frequencies drop faster
   in the black-solid and red-dashed curves.  This mainly originates from the dependence of $\phi_{0}$ on flux $\Phi_{e,s}$ ($\phi_{0}=0$ for the symmetric DC SQUID), because the effects of $\phi_s$ on the resonator modes' frequencies  are very small in the case of $|\phi_s|\ll1$.  The basics modes' frequencies  are shown in Fig.~\ref{fig2}(b),  so the effects of high-order modes can be neglected because of the large spectrum anharmonicities,  and the fundamental mode dominates the interaction with qubits and functions as a tunable coupler for Xmon qubits.


%

   Since $r_L, r_C\ll 1$,  the  second term in the right side of Eq.~(\ref{eq:5}) contributes
very little to the low-order modes, so the value of $k_0 \ell$ should be close to $\pi/2$.
 Next, we get an approximately analytical result for the  fundamental mode's  frequency,
 \begin{eqnarray}\label{eq:6}
 \omega^{(1)}_{c}\approx \frac{\omega^{(b)}_{c}}{1+ \frac{r_L}{2\sqrt{\cos^2\left(\frac{\pi\Phi_{e,s}}{\Phi_0}\right)+d^2\sin^2\left(\frac{\pi\Phi_{e,s}}{\Phi_0}\right)} \cos(\phi_s-\phi_0)}},
  \end{eqnarray}
  where  $\omega^{(b)}_{c}$ describes the  fundamental mode' frequency  of the bare resonator ($\Phi_{e,s}=0$).
   Since $\tan(k_n \ell/2)$ is a $\pi$-periodic function,
    the $n$-th resonator modes can be approximately written as $\omega^{(n)}_c=\omega^{(1)}_c+n\pi/\ell$.
The dispersion relation of the resonator coupler under the linear approximations is shown in Eqs.~(\ref{eq:5}) and~(\ref{eq:6}). After adding the nonlinear corrections induced by the SQUID,  the shift of $m$-th  energy level for $n$-th basic mode
 can be obtained as $\delta^{(n)}_m=-(6m^2+6m+3)\lambda_n E_{L,cav}$($m,n$ are integers),
 with $\lambda_n=\cos(k_n\ell)^2/[4(1+2k_n\ell/\sin(2k_n\ell))]$\cite{Sandberg1}.
 As shown in Fig.~\ref{fig2}(c), the anharmonicity $(\delta^{(1)}_2-\delta^{(1)}_1)$ for two lowest energy levels of fundamental mode
decreases when the magnetic flux passing through the SQUID loop  departs from half-integer flux quantum.
  According to the simulation result in Fig.~\ref{fig2}(d),
  an easy way of enhancing the fundamental mode's nonlinearity or anharmonicity
  is to reduce the transmission line resonator's capacitance.

Since the frequency tuning range for the resonator coupler is  about several hundred Megahertz, to switch off qubit-qubit coupling in the qubit-resonator large detuning regime,  the frequency of  bare resonator should be very important and it is determinated by the longitudinal length of resonator coupler. To effectively tune the frequency of the resonator coupler, the inductive energy of the DC SQUID should be much larger than that of the transmission line resonator, so  the large critical current Josephson junctions of the DC SQUID are needed. The transmission line resonator's inductance and capacitance per unit length are obtained from the experiment result \cite{IDA}. For the resonator coupler fabricated with a Niobium resonator and Aluminium SQUID, the detail parameters are: length $\ell= 4.87$ mm,
    inductance  per unit length 0.44 $\mu$H/m and capacitance  per unit length 0.16 nF/m,
   the inductive energy of transmission line resonator $E_{L,cav}/\hbar=[\hbar/(2e)^2]/(L_0 \ell)\approx 500$ GHz.
 The critical current of the DC SQUID  should be around $7.67\mu$A (for $r_L=0.02$),
   then the SQUID's maximal Josephson energy is $E^{(m)}_{Js}/\hbar=-\Phi_0 I_{cs}/(2\pi\hbar)\approx  2.4\times 10^4$ GHz.
Then the ratio of inductive energies for resonator and DC SQUID is $E_{L,cav}/E^{(m)}_{Js}\approx 1/100$. For the resonator coupler with above parameters should be easy to fabricate with current nano-fabrication techniques. Similar to the transmon-type coupler,  the quality factor of the resonator coupler will drop when an extra magnetic flux is applied to switch off/on qubit-qubit interaction\cite{Cleland}.   The resonator coupler with low quality factor  will induce dephasing of qubits, so it is important for the bare resonator to possess a relative high quality factor and the external magnetic flux to be farther from the half-integer quantum flux.


\section{Tunable resonator coupler}

\subsection{Tunable Qubit-Qubit  coupling}

\begin{figure}
\centering\includegraphics[bb=0 30 595 520, width=7.5cm, clip]{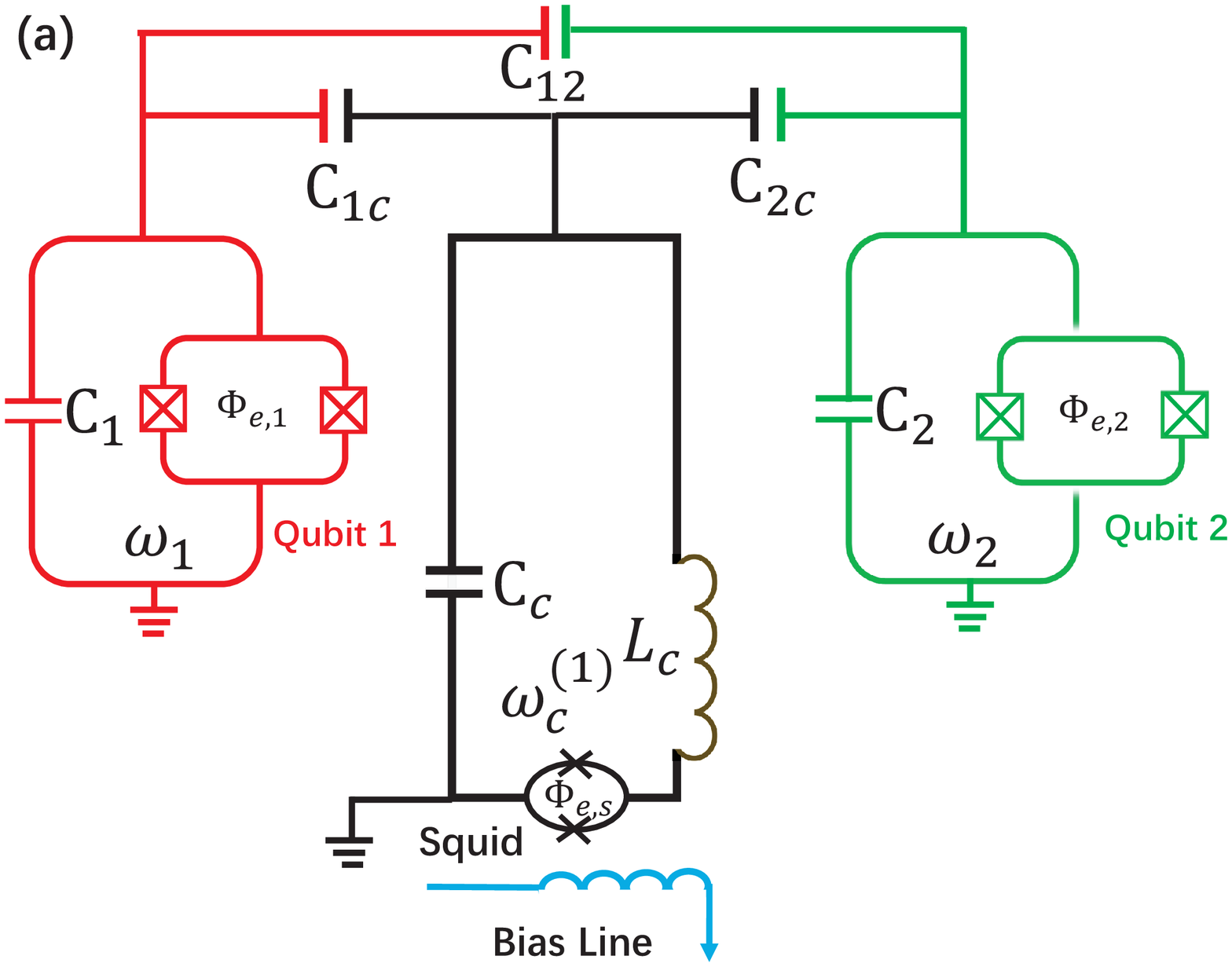}\\
\centering\includegraphics[bb=0 0 470 400, width=4.34cm, clip]{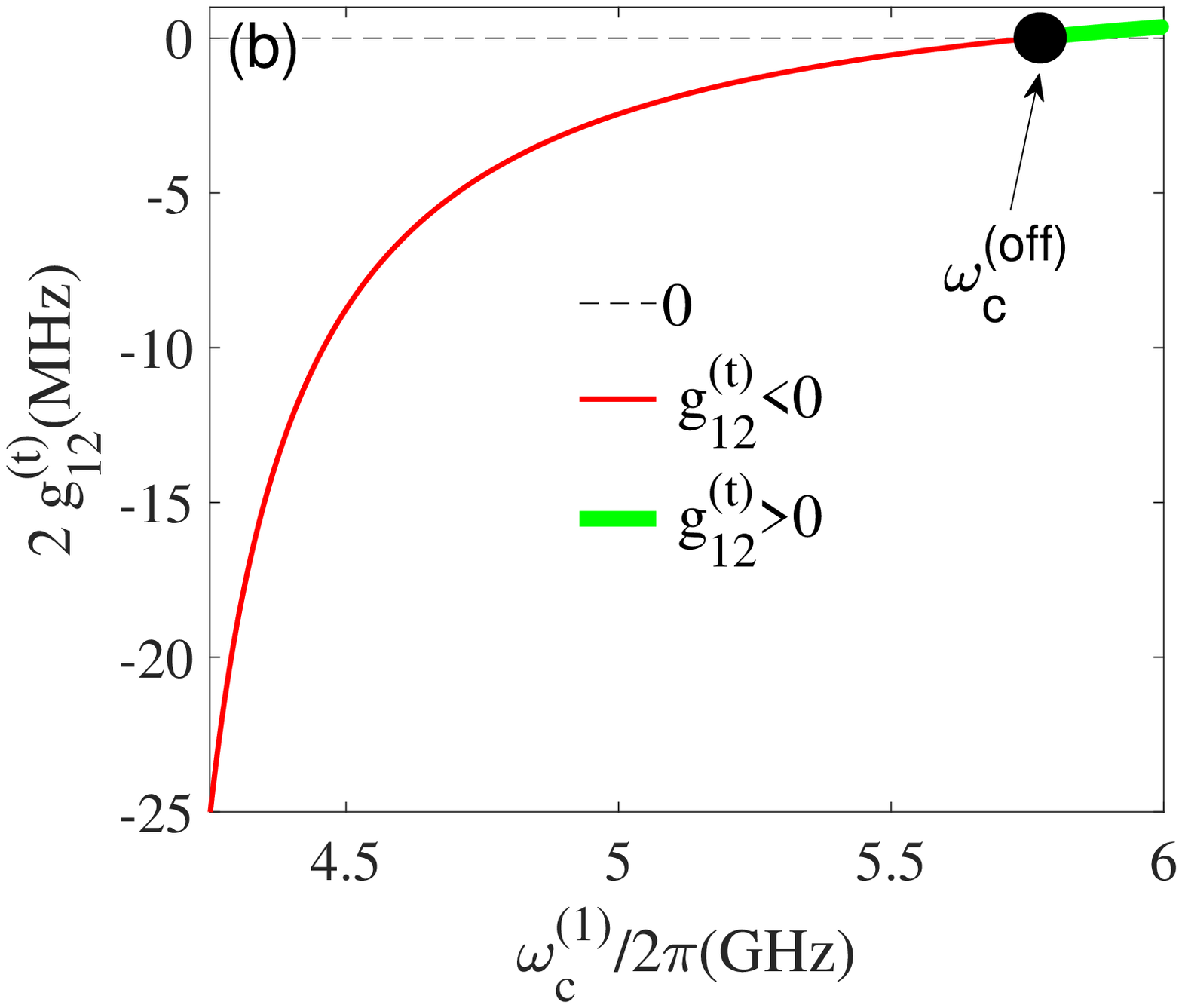}
\centering\includegraphics[bb=12 0 470 400, width=4.20cm, clip]{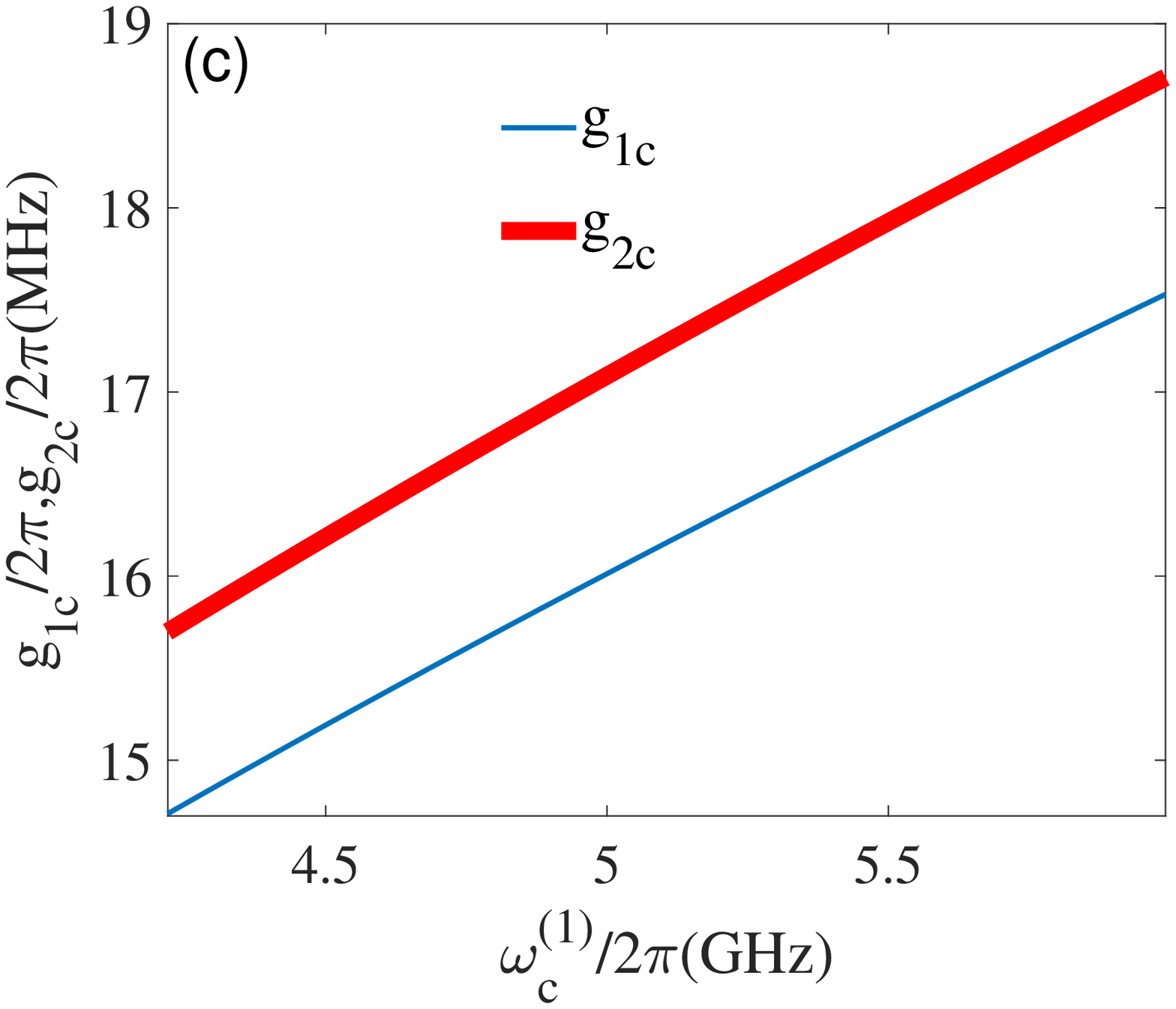}
\caption{(Color online)  Switching off/on the qubit-qubit coupling with the resonator coupler.
(a) Circuit Diagram of the three-body system. $C_{j}$ , $L_{j}$, $\omega_{j}$ , and  $\Phi_{e,j}$ ($j=1,2,c$) are the capacitances, inductances,  frequencies, and  magnetic fluxes  of Xmon qubits and resonator coupler, respectively.
 The quantity $C_{12}$ is the capacitance between two Xmon qubits,  $C_{1c}$ and  $C_{2c}$ are relative capacitances
 of  qubits 1 and 2 to  resonator coupler, respectively.
The  coupling strengths of (b) qubit-qubit  $g^{(t)}_{12}$ and (c)  qubit-resonator $g^{(1)}_{jc}$ ($j=1,2$)  changing with the fundamental mode's frequency $\omega^{(1)}_c$.
The other parameters are: $\omega_1/(2\pi)=4.0$ GHz, $\omega_2/(2\pi)=4.1$ GHz,
$r_L=0.02$, $r_C=0.1$, $d=0.1$, $g_{12}/(2\pi)=1.3 $ MHz, $C_c=0.78$ pF, $C_1=100$ fF,
$C_2=90$ fF, $C_{12}=0.06$ fF, $C_{1c}=1$ fF, and $C_{2c}=1$ fF.  }
\label{fig3}
\end{figure}


 Circuit diagram for the  three-body system is shown in Fig.~\ref{fig3}(a), the capacitive coupling between
any two of  qubit 1, qubit 2, and  resonator coupler depend on their frequencies and relative capacitances.
In convenience, here and afterwards we regard the T-shape quarter-wave resonator as a lumped-element model. The fundamental mode of resonator functions as a tunable coupler for the Xmon qubits, and its frequency,
 inductance,   and  capacitance are $\omega^{(1)}_c(\Phi_{e,s})$, $L_{c}(\Phi_{e,s})$, and $C_c$, respectively.
Besides, $C_{j}$ , $L_{j}$, $\omega_{j}$ ($j=1,2$)  are the capacitances, inductances and transition frequencies of qubits 1 and 2, respectively.
Additionally, $C_{12}$ is the relative capacitance between two qubits,
and $C_{jc}$  are the capacitances of qubits 1 and 2 relative to the resonator coupler, respectively.
If the direct effects of the DC SQUID on the Xmon qubits are neglected, the kinetic (charging) energy of the system can be written as
$T=\sum_{j=1,2}\big[(C_j/2) \dot{\phi}^2_j +(C_{jc}/2)(\dot{\phi}_j-\dot{\phi}_{c})^2\big]+(C_{c}/2) \dot{\phi}^{2}_{c}+
(C_{12}/2) (\dot{\phi}_1-\dot{\phi}_2)^2+\sum_{k=1,2}(C_{s_{k}}/2)\dot{\phi}^2_{s_k}$, and the
 potential (inductive) energy
 $U=-\sum_{j=1,2}E_{J_j}\cos(\phi_j)+E_{L,cav}(\phi_c)^2+E_{Js}\cos(\phi_s-\phi_0)$.
 The Lagrangian of the system can be obtained by $L=T-U$,
 after introducing  $q_{j}=\partial L/\partial \dot{\phi}_{j}$ ($j=1,2,c$),
  the Hamiltonian of three-body system can be written as $H_{ic}=\sum_{j=1,2,c}q_{j}\phi_{j}-L$.
  Expressed with device's parameters, we get
\begin{eqnarray}\label{eq:7}
H_{ic}&=&\sum^{2}_{j=1}\left[4 E_{C_j}(N_j)^2-E_{J_j}\cos(\phi_j)\right]\nonumber\\
&+&4 E_{C_c} (N_c)^2+E_{L,cav}(\phi_c)^2\nonumber\\
&+&4 E_{C_s}(N_s)^2-E_{Js}\cos(\phi_s-\phi_0)\nonumber\\
&+&\sum^{2}_{j=1}\left[\frac{8C_{jc}}{\sqrt{C_j C_c}}\sqrt{E_{C_j}E_{C_c}}(N_j N_c)\right]\\
&+&8\left(1+\frac{C_{1c} C_{2c}}{C_{12} C_{c}}\right)\frac{C_{12}}{\sqrt{C_1 C_2}}\sqrt{E_{C_1}E_{C_2}}(N_1 N_2).\nonumber
\end{eqnarray}
The  condition $C_c\gg  C_{1}, C_{2}\gg C_{1c}, C_{1c}\gg C_{12}$ have been used. Here $N_{j}$  are the number operators for charge quantization of Xmon qubits, and $\phi_j$ are the phase operators, where the subscripts $j=1,2$.
Their charging  and  Josephson energies are $E_{C_{j}}=e^2/2C_{j}$ and $E_{J_{j}}=-\Phi_0 I_{c_j}/2\pi$, respectively, and $I_{c_j}$ are the critical currents.  For Xmon qubits, $E_{J_j}/E_{C_j} \gg 1$, then we can approximately obtain $-E_{J_{j}}\cos(\phi_{j})\approx - E_{J_{j}}+(E_{J_{j}}/2)\phi^2_{j}-(E_{J_{j}}/24)\phi^4_{j}+\cdot\cdot\cdot$, where the subscripts $j=1,2$.
If we define $\phi_j=(2E_{C_j}/E_{J_j})^{1/4}(c_j+c^{\dagger}_j)$ and $N_j=i[E_{J_j}/(32E_{C_j})]^{1/4}(c_j-c^{\dagger}_j)$,
 and the Hamiltonian of Xmon qubit can be written as $H_{q_j}=\sqrt{8 E_{C_j} E_{J_j}}(c^{\dagger}_j c_j+1/2)-(E_{C_j}/12)(c^{\dagger}_j+c_j)^4-E_{J_j}$,
 and $c^{\dagger}_j$($c_j$) are the creation (annihilation) operators of excited modes, where the subscripts $j=1,2$.
   The Xmon qubit's lowest transition energy  level is
   $\omega_j=(\sqrt{8 E_{C_j} E_{J_j}}-E_{C_j})/\hbar$,
with the strength of anharmonicities $\alpha_j=-E_{C_j}/\hbar$.

For the  resonator coupler, $\phi_{c,s}$  are the superconducting phase operators of transmission line resonator and DC SQUID, respectively, while $N_{c,s}$  are their charge  number operators. The $E_{C_{s}}=e^2/2(C_{s_{1}}+C_{s_{2}})$ and $E_{Js}$ are  the total charging energy and Josephson energy of the DC SQUID, respectively.   The inductive energy  of the DC SQUID is much larger than that of the transmission line resonator, while the relation for charging energies is on the contrary, so the fundamental mode's frequency is determined by the inductive energy of the DC SQUID and charge energy of the transmission line resonator.
 The creation and annihilation operators of photons of the coupler  resonator can be introduced by $N_c=i[E_{Js}/(32E_{C_c})]^{1/4}(a_1-a^{\dagger}_1)$ and $\phi_c=(2E_{C_c}/E_{Js})^{1/4}(a_1+a^{\dagger}_1)$,
 and the frequency of fundamental mode for bare resonator ($\Phi_{e,s}=0$) is $\omega^{(b)}_c=\sqrt{8E_{C_c} E_{Js}}/\hbar$. If an external magnetic flux is applied,   the fundamental mode's frequency can be obtained by solving Eq.~(\ref{eq:5}), then
 the effective Hamiltonian of tunable coupler becomes $H_{tc}=\hbar(\omega^{(1)}_{c}+\delta^{(1)}_m)a^{\dagger}_{1}a_{1}$, while the $\delta^{(1)}_m$ is the nonlinear energy level shift of $m$-photon state for the fundamental mode.


With the creation and annihilation operators,  the Hamiltonian for the direct qubit-qubit  interaction is $H_{12}= -\hbar g_{12}(c_{1}-c^{\dagger}_{1})(c_2-c^{\dagger}_2)$,
 while the interaction Hamiltonians between the resonator coupler and qubits are $H_{jc}= -\hbar g^{(1)}_{jc}(a_{1}-a^{\dagger}_{1})(c_j-c^{\dagger}_j)$, where the subscripts $j=1,2$.
Expressed with the device's parameters, then the direct qubit-qubit coupling strength becomes $g_{12}=[\left(C_{12}+C_{1c} C_{2c}/C_c\right)/(2\sqrt{C_1 C_2})]\sqrt{\omega_1\omega_2}$, and the qubit-resonator coupling strengths are $g^{(1)}_{jc}=[C_{jc}/(2\sqrt{C_j C_c})]\sqrt{\omega_j \omega^{(1)}_c}$ which also depends on the coupler's frequency $\omega^{(1)}_c$
as shown in Fig.~\ref{fig3}(c). To see clearer the indirect qubit-qubit interaction induced by the resonator coupler,
 we apply the Schrieffer-Wolf transformation, that is,
  $U=\exp\big\{ \sum_{j=1,2}[(g^{(1)}_{jc}/\Delta_j)(c^{\dagger}_{j}a_{1}-c_{j}a^{\dagger}_{1})-(g^{(1)}_{jc}/\Lambda_j)(c^{\dagger}_{j}a^{\dagger}_{1}-c_{j}a_{1})]\big\}$,
  with $\Delta_j=\omega_j-\omega^{(1)}_c$ and  $\Lambda_j=\omega_j+\omega^{(1)}_c$ ($j=1,2$).
  Then the  resonator coupler is decoupled with Xmon qubits, and the decoupled Hamiltonian is
   $H_{d}=\sum_{j=1,2}\left[\tilde{\omega}_{j}c^{\dagger}_{j}c_{j}+\frac{\tilde{\alpha}_{j}}{2}c^{\dagger}_{j}c^{\dagger}_{j}c_{j}c_{j}\right]+g^{(t)}_{12}(c^{\dagger}_{1}c_2+c_{1}c^{\dagger}_2)$,
where  $\tilde{\omega}_j \approx  \omega_j+[g^{(1)}_{jc}]^2(1/\Delta_j-1/\Lambda_j)$,
$\tilde{\alpha}_j \approx \alpha_j$, where the subscripts $j=1,2$.
The effective coupling strength between the two qubits is
 \begin{eqnarray}\label{eq:8}
g^{(t)}_{12}&\approx & \frac{1}{2}\left(1+\frac{C_{1c} C_{2c}}{C_c C_{12}}\right)\frac{C_{12}}{\sqrt{C_{1} C_{2}} }\sqrt{\omega_1 \omega_2}\\
 & +&\frac{\omega^{(1)}_c}{8}\left[\frac{1}{\Delta_1}+\frac{1}{\Delta_2}-\frac{1}{\Lambda_1}-\frac{1}{\Lambda_2} \right] \frac{C_{1c} C_{2c}}{C_c\sqrt{C_1 C_2} }\sqrt{\omega_1 \omega_2}.\nonumber
\end{eqnarray}
The first line in Eq.~(\ref{eq:8}) describes the direct capacitive qubit-qubit coupling,
and the second line corresponds to the indirect qubit-qubit interaction induced by the  resonator coupler. As shown in Fig.~\ref{fig2}, the frequency $\omega^{(1)}_c$ can be effectively tuned by the external magnetic flux,  so the fundamental mode of T-shape quarter-wave resonator can function as a tunable coupler for Xmon qubits.
The  qubit-qubit coupling is totally switched off ($g^{(t)}_{12}=0$) at a certain frequency $ \omega^{(off)}_c =4[1+C_{12} C_{c}/(C_{1c} C_{2c})]/(1/\Lambda_1+1/\Lambda_2-1/\Delta_1-1/\Delta_2)$ as show in Fig.~\ref{fig3}(b).
  As shown in Eq.~(\ref{eq:5}), the  indirect qubit-qubit coupling is proportional to the qubit-resonator coupling strength which is greatly affected by resonator coupler's parameters including the frequency $\omega^{(1)}_c$ (see Fig.~\ref{fig3}(c)), capacitance $C_c$, and the relative capacitances $C_{jc}$. And the direct qubit-qubit coupling is affected by  relative capacitance between two qubits which can be weakened by enhancing the transverse broaden part of the resonator coupler in the open ends.
 So the T-shape quarter-wave resonator can help to realize a strong enough indirect coupling to switch off the qubit-qubit interaction and also a weak direct qubit-qubit interaction, which is an important way to suppress the ZZ crosstalk (as will be discussed in the follwing Sections).


\subsection{ZZ crosstalk}

The Parasitic crosstalk  between neighbour superconducting qubits is a leading limitation
for quantum gates, and it is greatly affected by the anharmonicities of  the qubits and resonator coupler.
With the devices consisting of two couplers, some experiments finally cancel the qubit-qubit coupling and ZZ crosstalk\cite{Kandala,Mundada}, but this will enhance complexities for the design, fabrication, and measurement processes, and some extra energy levels should be introduced. By assuming the opposite anharmonicities of two qubits, some research group proposed to totally cancel the ZZ crosstalk \cite{Zhao}, but this should be very difficult in experiment. In this paper, we suppress  the  ZZ crosstalk by reducing the direct  qubit-qubit coupling with the transversely broaden part of the T-shape quarter-wave resonator.

\begin{figure}
\centering\includegraphics[bb=0 0 580 540, width=7.2cm, clip]{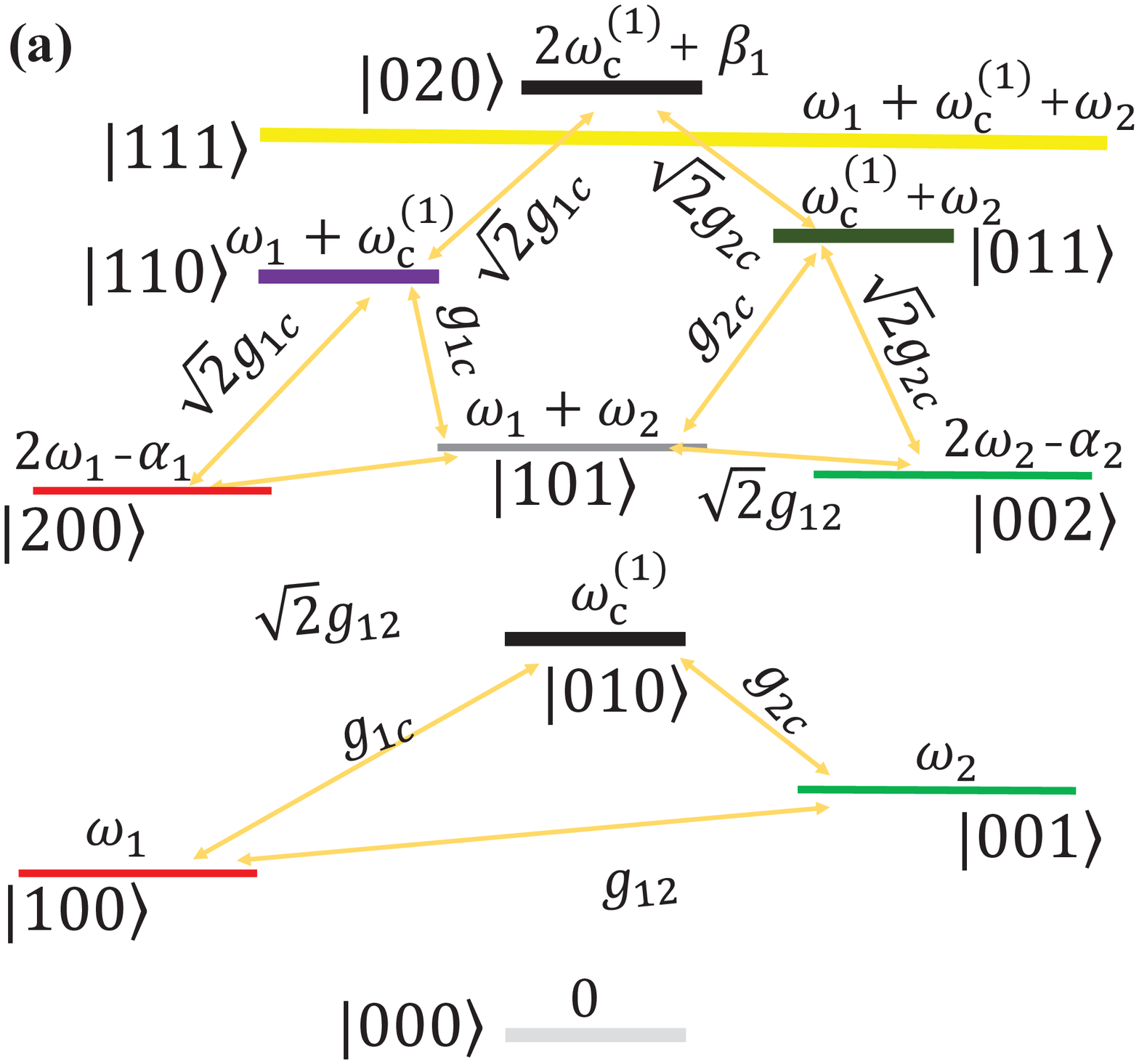}\\
\centering\includegraphics[bb=0 0 515 445, width=4.28cm, clip]{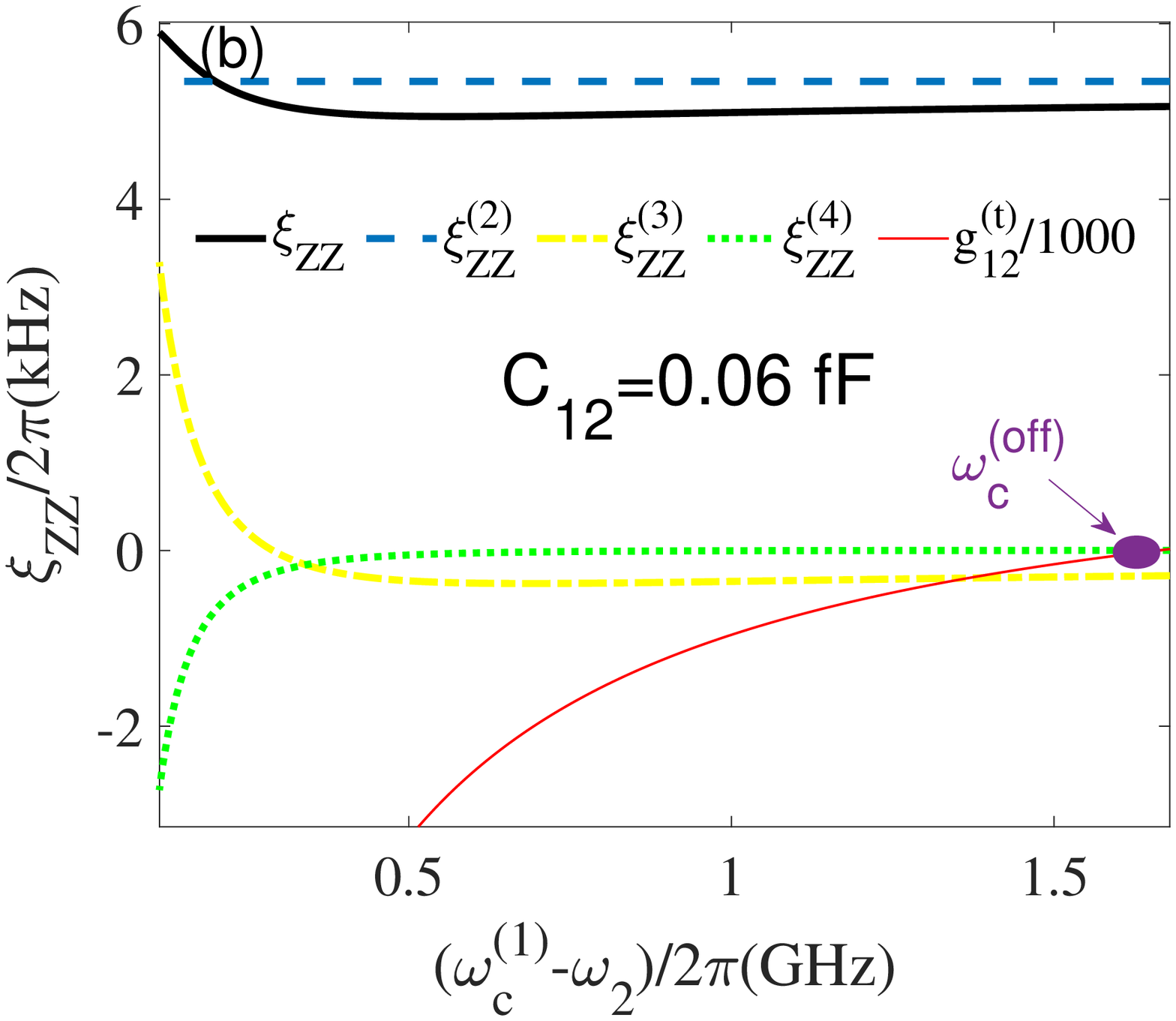}
\centering\includegraphics[bb=0 0 515 445, width=4.28cm, clip]{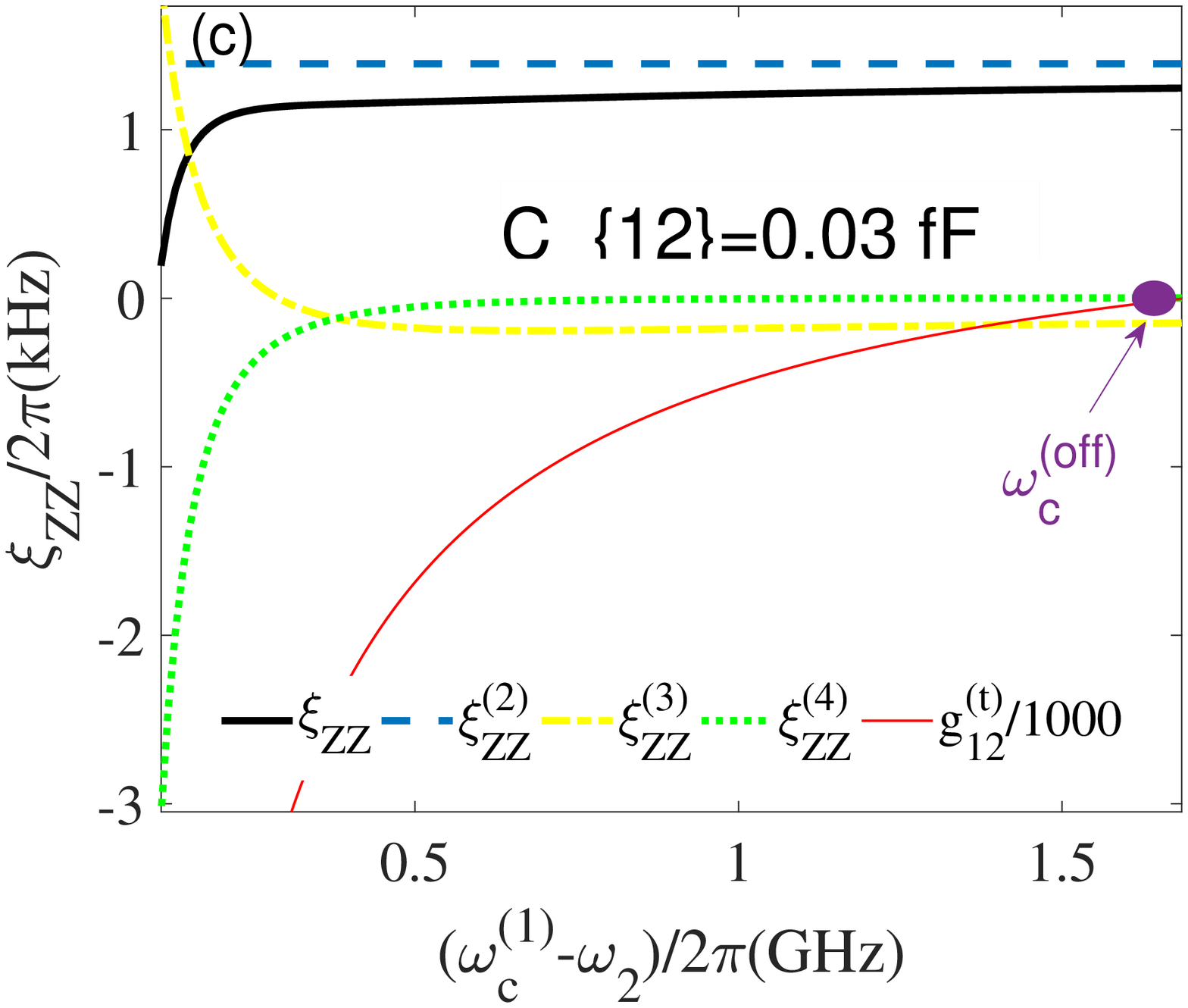}
\caption{(Color online) Residual ZZ crosstalk. (a) Energy level structure
   of the three-body system in the idle state. The ZZ crosstalk $\xi_{ZZ}$
  for the capacitance (b) $C_{12}=0.06$ fF and (c)  $C_{12}=0.03$ fF.
     The parameters of the DC SQUID are the same as the  black-solid curve
     in Fig.~\ref{fig2}(a), while the parameters of the qubits and transmission line resonator
       are the same as in Fig.~\ref{fig3}(b).}
\label{fig4}
\end{figure}

Figure~\ref{fig4}(a) describes the energy level structure of the three-body system with
qubits in the idle states, and the  fundamental mode of resonator coupler is dispersive coupling with Xmon qubits.
In the processes of two qubit quantum gates, some states will experience near resonant energy exchanges, such as $|100\rangle\leftrightarrow |001\rangle$ and $|101\rangle \leftrightarrow|200\rangle$ for the CZ (Controlled-Z) gates. According to the discussions in above Sections,
after neglecting some virtual transition processes,  the Hamiltonian of the three-body system becomes
\begin{eqnarray}\label{eq:9}
H_{rwa}&=&\hbar\left(\omega^{(1)}_{c}+\delta^{(1)}_m\right)a^{\dagger}_{1}a_{1}\nonumber\\
&+&\sum^2_{j=1}\left[\hbar\omega_{j}c^{\dagger}_{j}c_{j}+\frac{\hbar\alpha_{j}}{2}c^{\dagger}_{j}c^{\dagger}_{j}c_{j}c_{j}\right]\\
& +&\sum^2_{j=1}\hbar g^{(1)}_{jc}\left(c_{j}a^{\dagger}_1+c^{\dagger}_{j}a_1\right) +\hbar g_{12}\left(c_{1}c^{\dagger}_2+c^{\dagger}_{1}c_2\right),\nonumber
\end{eqnarray}
 the rotating wave approximation has been used above. The $\delta^{(1)}_m$ is the nonlinear energy level shift of the $\textit{m}$-photon states for the resonator's fundamental mode, both $\omega^{(1)}_{c}$ and $\delta^{(1)}_m$ can be tuned by the external magnetic flux passing through the SQUID loop.

If the frequencies of qubits 1 and 2  are detuning, the residual ZZ crosstalk for the CZ gate
can be defined as  $\xi_{ZZ}=\omega_{11}-\omega_{01}-\omega_{10}$\cite{Sun,Sete,Sung}.
According to the numerical calculation in section III,  we choose an anharmonicity value $\delta^{(1)}_2-\delta^{(1)}_1=-50$ MHz  for  two lowest energy levels of the fundamental mode, the frequency $\omega^{(1)}_c/2\pi$ is close to $4.5$ GHz according to the curve in Fig.~\ref{fig2}(c).
The total ZZ crosstalk  can be written as $\xi_{ZZ}=\xi^{(2)}_{ZZ}+\xi^{(3)}_{ZZ}+\xi^{(4)}_{ZZ}$,
up to fourth order small quantity \cite{Pople,Sung,Sun,Zhu}, we get
\begin{eqnarray}\label{eq:10}
\xi^{(2)}_{ZZ}&=&\frac{2 (g_{12})^2(\alpha_1+\alpha_2)}{(\Delta_{12}+\alpha_1)(\Delta_{12}-\alpha_2)},\\
\xi^{(3)}_{ZZ}&=&2 g_{12}g^{(1)}_{1c}g^{(1)}_{2c}\bigg[\frac{1}{\Delta_{1c}}\left(\frac{2}{\Delta_{12}-\alpha_2}-\frac{1}{\Delta_{12}}\right),\nonumber\\
& &-\frac{1}{\Delta_{2}}\left(\frac{2}{\Delta_{12}+\alpha_2}-\frac{1}{\Delta_{12}}\right)\bigg],\\
\xi^{(4)}_{ZZ}&=&\frac{2[g^{(1)}_{1c}g^{(1)}_{2c}]^2}{\Delta_{1}+\Delta_{2}+\delta^{(1)}_1-\delta^{(1)}_2}\left(\frac{1}{\Delta_{1c}}+\frac{1}{\Delta_{2c}}\right)^2\nonumber\\
      & &+\frac{[g^{(1)}_{1c}g^{(1)}_{2c}]^2}{\Delta^2_{1}} \left(\frac{2}{\Delta_{12}-\alpha_2}-\frac{1}{\Delta_{12}}-\frac{1}{\Delta_{2}}\right)\nonumber\\
      & &-\frac{[g^{(1)}_{1c}g^{(1)}_{2c}]^2}{\Delta^2_{2c}} \left(\frac{2}{\Delta_{12}+\alpha_1}-\frac{1}{\Delta_{12}}+\frac{1}{\Delta_{1}}\right).
\end{eqnarray}
Here $\Delta_{12}=\omega_{1}-\omega_{2} $ is the detuning between two qubits,
and the $\omega_{1}$ is fixed here.  As shown in Figs.~\ref{fig4}(b) and \ref{fig4}(c), $|\xi^{(2)}_{ZZ}|\gg |\xi^{(3)}_{ZZ}|\gg |\xi^{(4)}_{ZZ}|$, the second order term $\xi^{(2)}_{ZZ}$  dominates  the ZZ crosstalk in the qubit-resonator dispersive regime,\cite{Sun,Mundada,Zhu}.  The black-solid  curve in Fig.~\ref{fig4}(b) shows that  the total ZZ crosstalk
 is in the magnitude order of  several kilohertz, which is possible for realizing high fidelity quantum gates.

   The $\xi^{(2)}_{ZZ}$ in Eq.(10) is proportional to the square of direct qubit-qubit coupling ($g^2_{12}$) and the total anharmonicity  $(\eta_1+\eta_2)$ of two Xmon qubits, while the third-order term $\xi^{(3)}_{ZZ}$  in Eq.(11) and fourth-order term $\xi^{(4)}_{ZZ}$  in Eq.(12) mainly depend
on the qubit-resonator coupling strengths $g^{(1)}_{jc}$,  where the subscripts $j=1,2$.
If we  increase the separation  between
two Xmon qubits with the help of the T-shape resonator coupler, then the relative capacitance $C_{12}$ and direct qubit-qubit coupling strength $g_{12}$ become smaller,  then the ZZ crosstalk can be effectively suppressed as shown in Fig.~\ref{fig4}(c).



%
%

\subsection{States leakages}

In this paper, the switching off frequency for qubit-qubit coupling lies in the strong dispersive regime. Then the weak negative anharmonicity of the fundamental mode [see Fig.~\ref{fig2}(c)] should push the state $|020\rangle$ farther from the computational states,
and this might reduce the leakage probability to the  resonator coupler's double-photon states.
The anharmonicity of the Xmon qubit is about several hundred Megahertz,
   which is much larger than qubit-resonator coupling strength.  By introducing the bright and dark modes,
   the dynamic evolution in the CZ  gate can be described by the following Hamiltonian\cite{Chen,Sung}
  \begin{eqnarray}\label{eq:11}
H^{CZ}_{|100\rangle\leftrightarrow|010\rangle}&=&\left(
           \begin{array}{cc}
             \omega_1 & g^{(1)}_{1c} \\
             g^{(1)}_{1c} & \omega^{(1)}_c \\
           \end{array}
         \right), \\
 H^{CZ}_{|101\rangle\leftrightarrow|B_{CZ}\rangle}&=&\left(
 \begin{array}{cc}
 \omega_1+\omega_2 & g^{(1)}_{1c} \\
  g^{(1)}_{1c} & \omega^{(1)}_c+\omega_2 \\
           \end{array}
         \right).
 \end{eqnarray}

   \begin{figure}
\centering\includegraphics[bb=0 0 543 435, width=4.25cm, clip]{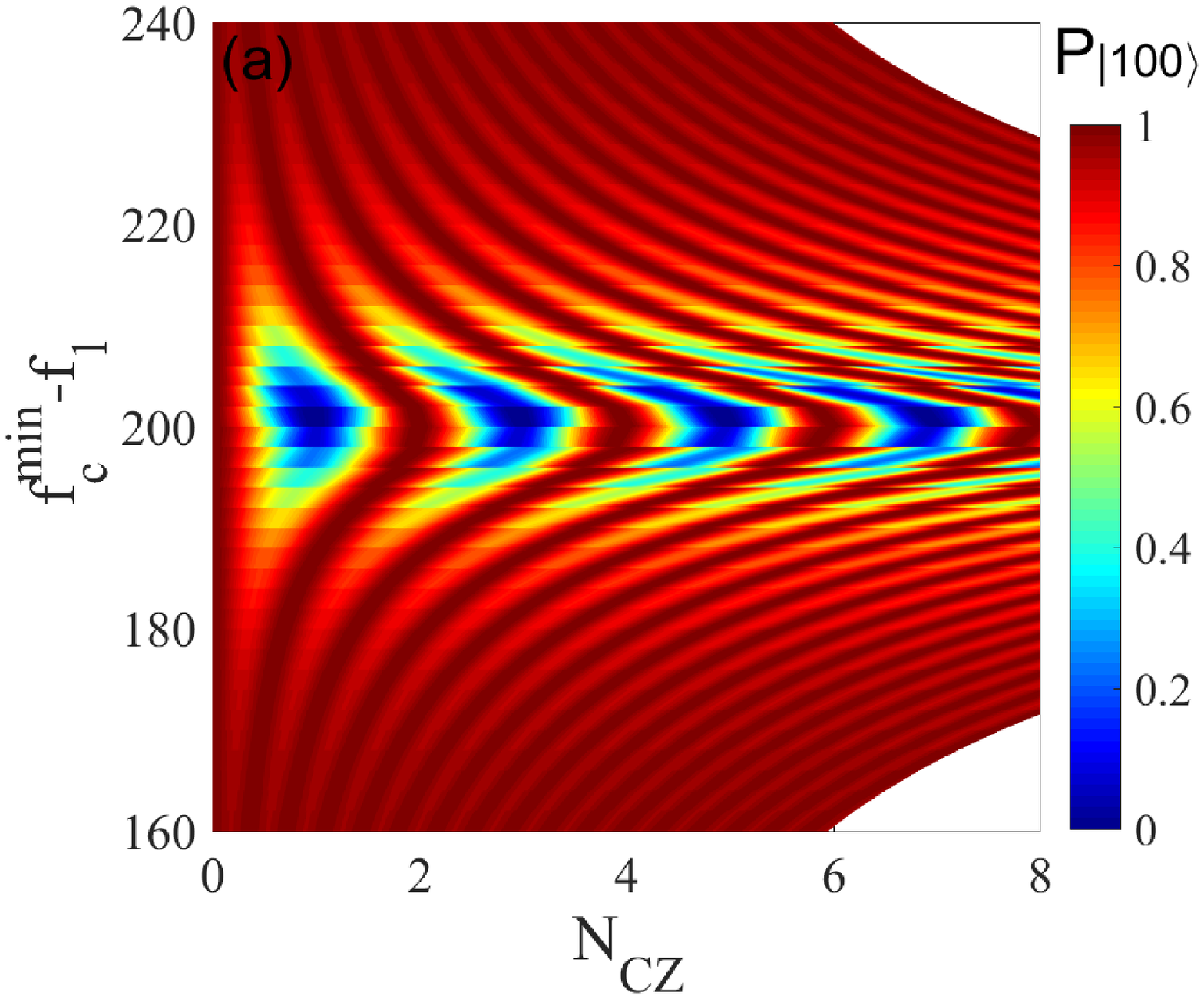}
\centering\includegraphics[bb=0 0 543 435, width=4.25cm, clip]{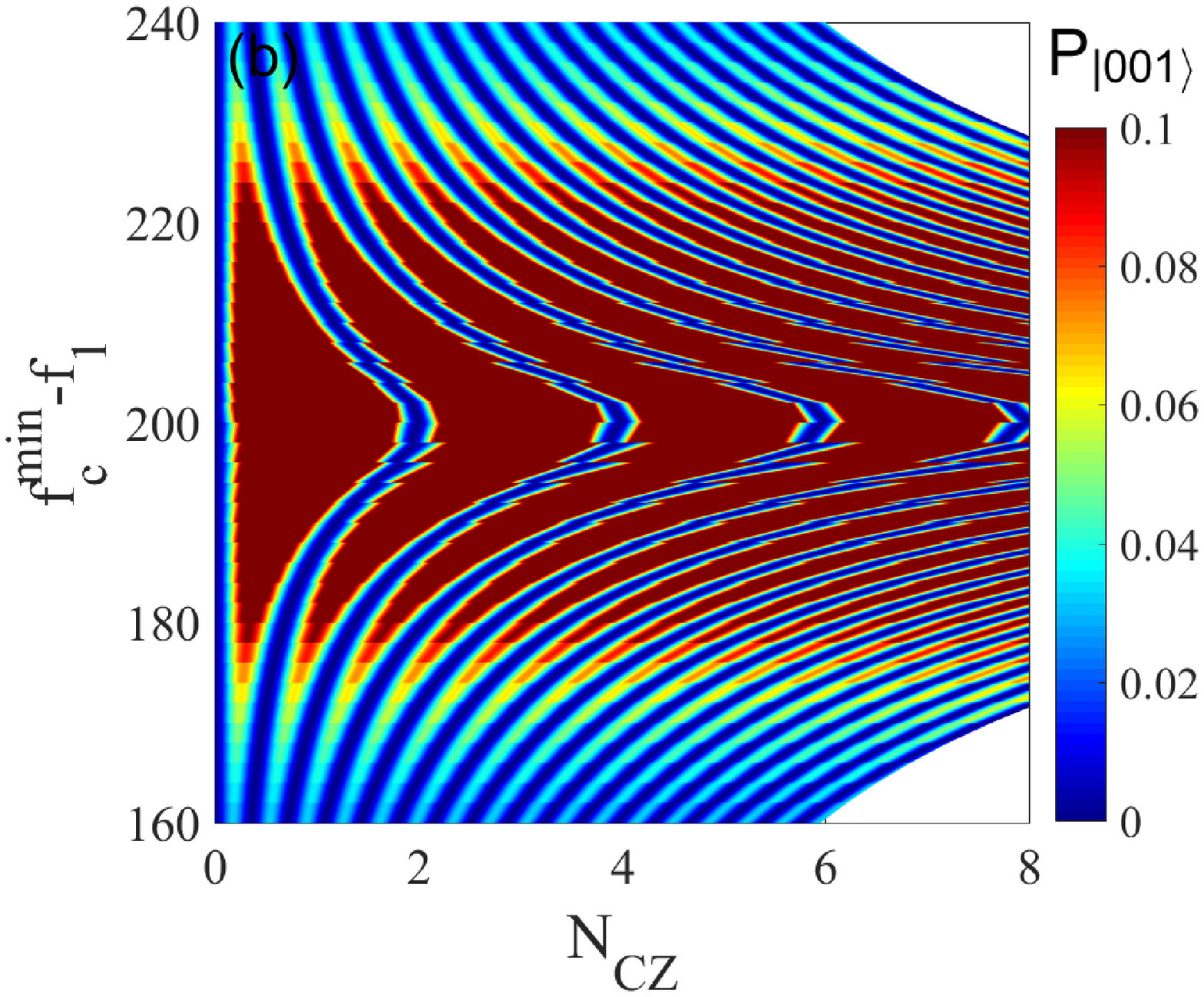}\\
\centering\includegraphics[bb=0 0 543 435, width=4.25cm, clip]{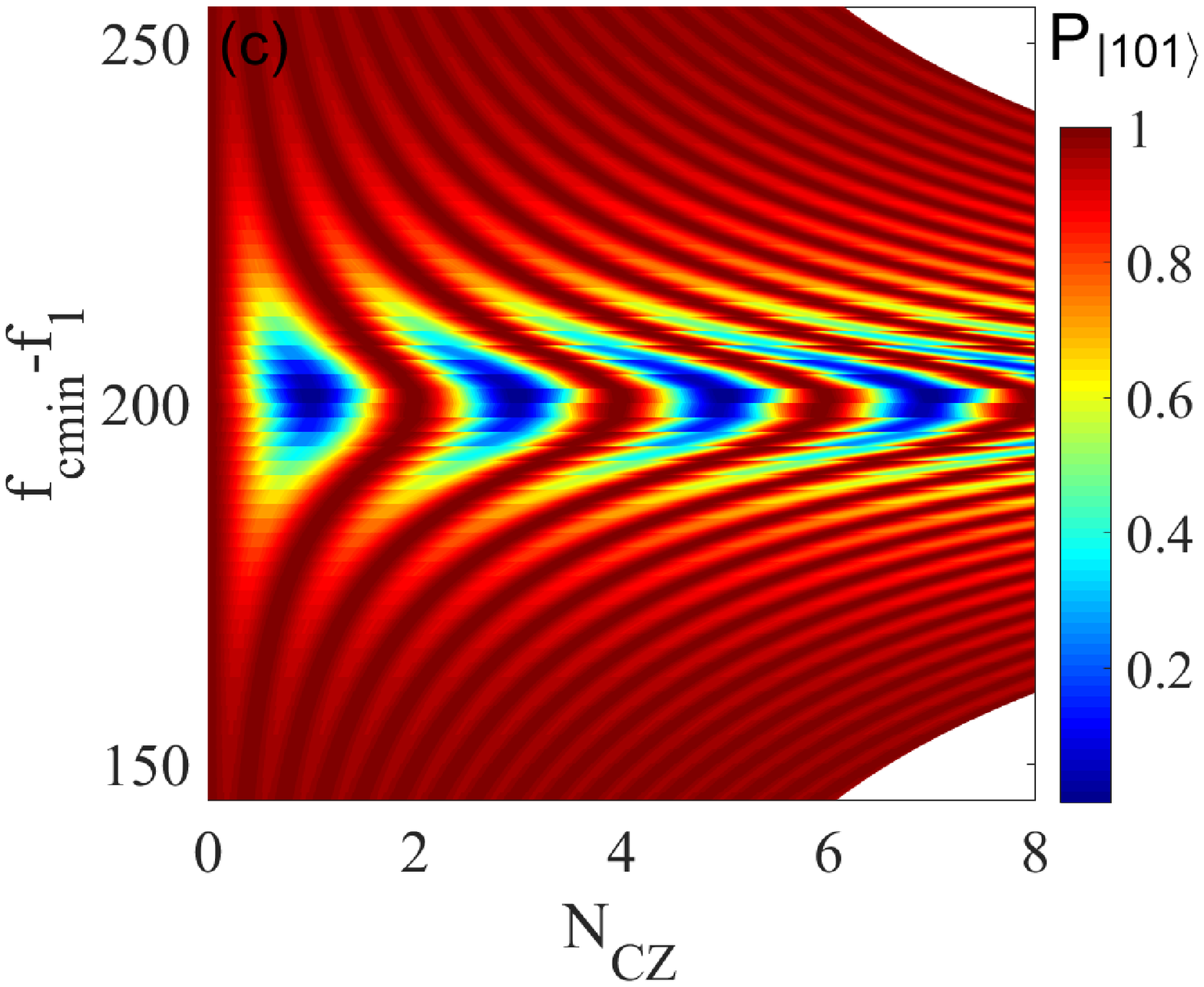}
\centering\includegraphics[bb=0 0 543 435, width=4.25cm, clip]{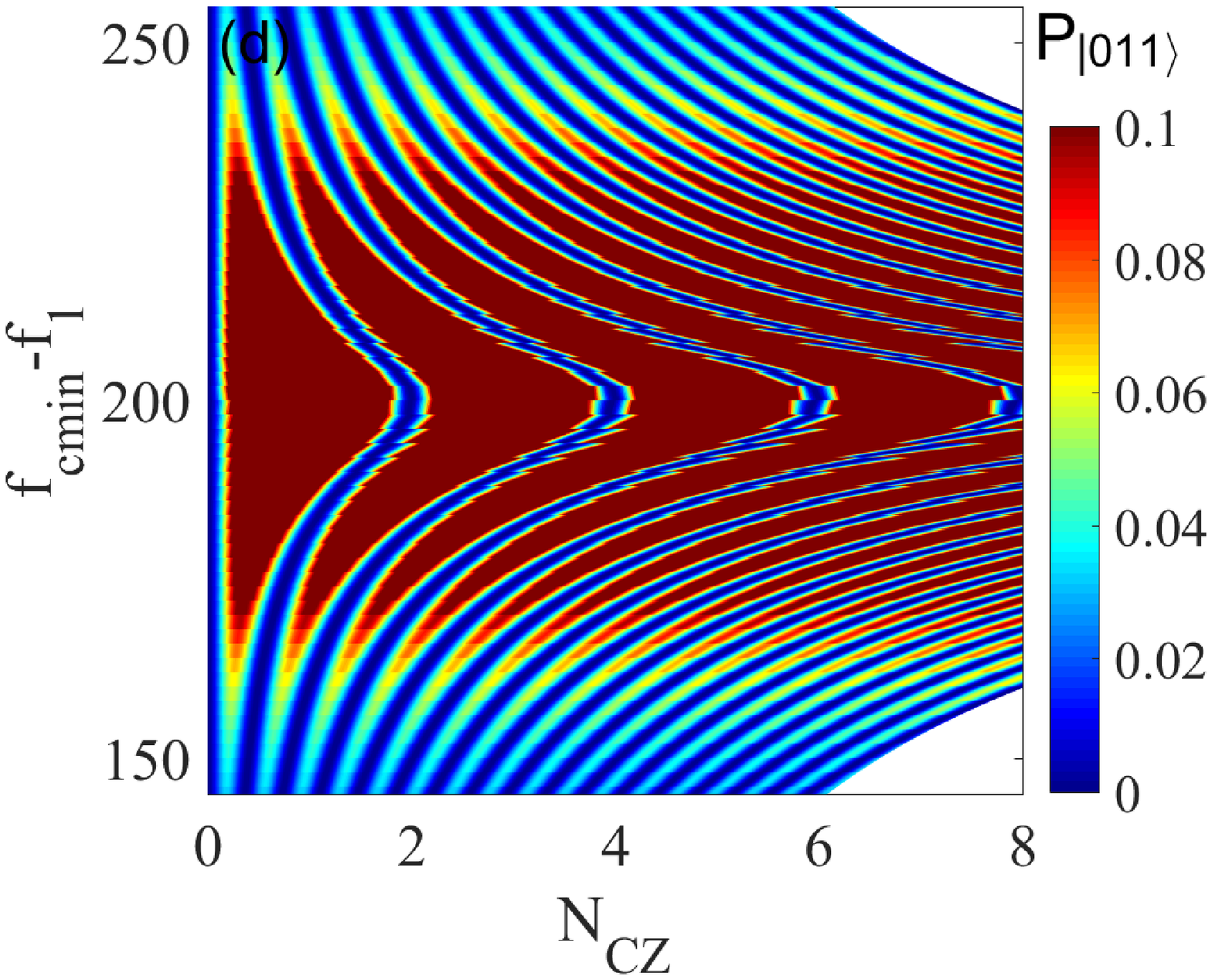}
\caption{(Color online) State leakage to the resonator coupler during CZ gate.
Populations of states (a) $|100\rangle$, (b) $|010\rangle$ , (c) $|101\rangle$, and (d) $|011\rangle$
 changing with the pulse amplitude and CZ gate number $N_{CZ}$. The prepared state is $|100\rangle$  for
  (a) and (b), but $|101\rangle$ for (c) and (d).
   The parameters of the resonator coupler are the same as in Fig.~\ref{fig3}(b). }
\label{fig5}
\end{figure}

 During the CZ gate,  for a prepared state $|100\rangle$,
  the computational  states are  $|100\rangle$ and  $|101\rangle$.
  The dominant leakage channel appears within the near resonant energy exchange between state $|100\rangle$
    and non-computational state  $|010\rangle$, and the state leakage dynamic
    resembles  the off-resonant Rabi oscillation which can be described
    by the two-level Hamiltonian $H^{CZ}_{|100\rangle\leftrightarrow|010\rangle}$ in Eq.(13)\cite{Chen,Sung}.
    The periodic oscillation of $P_{|010\rangle}$ in  Fig.~\ref{fig5}(b) describes the leakage probabilities
      to the state $|010\rangle$  during the CZ gate, which can be suppressed by tuning the wave shape
       of microwave pulses.

    If the initially prepared state is  $|101\rangle$,
  where the  dominant  leakage states appear within the non-computational states $|011\rangle$ and $|200\rangle$,
   and the leakage dynamics is described by the Hamiltonian  of the three-level system\cite{Sung}.
  To simplify the dynamic process,
 the bright mode  hybridization $|B_{CZ}\rangle=\cos(\vartheta)|011\rangle+\sin(\vartheta)|200\rangle$
 and dark mode hybridization $|D\rangle=\sin(\vartheta)|011\rangle+\cos(\vartheta)|200\rangle$ can be introduced,
 with $\tan\vartheta=\sqrt{2}g_{12}/g^{(1)}_{1c}$.
 In the case of $g^{(1)}_{1c}\gg g_{12}$,
  we have $\vartheta\approx 0$, the dark mode is decoupled from states $|101\rangle$ and  $|B_{CZ}\rangle$,
   the bright mode  $|B_{CZ}\rangle$ reduces to $|011\rangle$
   with a transition frequency $\omega_B=\omega^{(1)}_c+\omega_2$,
    and the coupling strength between states $|101\rangle$ and $|B_{CZ}\rangle$
    is approximately equal to $g^{(1)}_{1c}$.
    Then the leakage dynamic process can be described by a reduced two-level Hamiltonian
    $H^{CZ}_{|101\rangle\leftrightarrow|B_{CZ}\rangle}$  in Eq.(14) and the leakage probabilities $P_{|B_{CZ}\rangle}$  to bright mode $|B_{CZ}\rangle$ are described by the  periodic oscillation in Fig.~\ref{fig5}(d).

\section{Conclusions}\label{conclusion}

In conclusion, we have proposed a tunable coupler scheme  for superconducting
Xmon qubits based on a T-shape quarter-wave resonator. In experimentally accessible parameter regime, we studied the effective qubit-qubit coupling, residual
ZZ crosstalk, and state leakages. Our  tunable coupler scheme is easy for the nano-fabrication and
measurement, and  less magnetic flux noises will be created to the Xmon qubits. The main limitation for our scheme is to keep the resonator coupler with a relative high quality factor during the  frequency tuning by the external magnetic flux.  The open ends of the T-shape quarter-wave resonator could  help to suppress the ZZ crosstalk and realize high fidelity quantum gates.

\section{ACKNOWLEDGMENTS}

H. W. thanks valuable suggestions from Xiu Gu and Yarui Zheng, and support from Inspur
Academy of science and technology. Y.J.Z. is supported by Beijing Natural Science Foundation
under Grant No.  4222064 and National Natural Science Foundation of China under Grant No. 11904013.
X.-W.X. is supported by the National Natural Science Foundation of China
under Grant No.~12064010, and Natural Science Foundation of Hunan Province of China under Grant No.~2021JJ20036.


\begin{thebibliography}{99}


\bibitem{Chen} Y. Chen, C. Neill, P. Roushan, N. Leung, M. Fang, R. Barends, J. Kelly, B. Campbell, Z. Chen, B. Chiaro,
 A. Dunsworth, E. Jeffrey, A. Megrant, J. Y. Mutus, P. J. J. \'{O}Malley, C. M. Quintana, D. Sank,
 A. Vainsencher, J. Wenner, T. C. White, Michael R. Geller, A. N. Cleland, and
J. M. Martinis, qubit Architecture with High Coherence and Fast Tunable Coupling, Phys. Rev. Lett. \textbf{113}, 220502
(2014).


\bibitem{Yan}F. Yan, P. Krantz, Y. Sung, M. Kjaergaard, D. L. Campbell,
T. P. Orlando, S. Gustavsson, and W. D. Oliver, Tunable Coupling Scheme for
 Implementing High-Fidelity Two-qubit Gates, Phys. Rev. Applied \textbf{10}, 054062 (2018).


\bibitem{Sun} X. Li, T. Cai, H. Yan, Z. Wang, X. Pan, Y. Ma, W. Cai, J. Han,
 Z. Hua, X. Han, Y. Wu, H. Zhang, H. Wang, Yipu Song, Luming Duan, and Luyan Sun,
 Tunable Coupler for Realizing a Controlled-Phase Gate with Dynamically Decoupled Regime in a Superconducting Circuit,
Phys. Rev. Applied \textbf{14}, 024070 (2020).



\bibitem{Allman} M. S. Allman, F. Altomare, J. D. Whittaker, K. Cicak, D. Li, A. Sirois,
 J. Strong, J. D. Teufel, and R. W. Simmonds, rf-SQUID-Mediated Coherent Tunable Coupling
  between a Superconducting Phase qubit and a Lumped-Element Resonator, Phys. Rev. Lett. \textbf{104}, 177004 (2010).

\bibitem{Ming} Ming Gong, Shiyu Wang, Chen Zha, Ming-Cheng Chen, He-Liang Huang, Yulin Wu,
 Qingling Zhu, Youwei Zhao, Shaowei Li, Shaojun Guo, Haoran Qian, Yangsen Ye, Fusheng Chen,
 Chong Ying, Jiale Yu, Daojin Fan, Dachao Wu, Hong Su, Hui Deng, Hao Rong, Kaili Zhang, Sirui Cao,
  Jin Lin, Yu Xu, Lihua Sun, Cheng Guo, Na Li, Futian Liang, V. M. Bastidas, Kae Nemoto, W. J. Munro,
   Yong-Heng Huo, Chao-Yang Lu, Cheng-Zhi Peng, Xiaobo Zhu, Jian-Wei Pan,
   Quantum walks on a programmable two-dimensional 62-qubit superconducting processor, Science \textbf{372}, 948 (2021).

\bibitem{IBM} IBM-Q-Team, IBM-Q-53 Rochester backend specification
v1.2.0, 2020.

\bibitem{Kandala} A. Kandala, K. X. Wei, S. Srinivasan, E. Magesan, S. Carnevale, G. A. Keefe,
 D. Klaus, O. Dial, and D. C. McKay, Demonstration of a High-Fidelity cnot Gate for
 Fixed-Frequency Transmons with Engineered ZZ Suppression, Phys. Rev. Lett. \textbf{127}, 130501 (2021).

\bibitem{Sete} E. A. Sete, A. Q. Chen, R. Manenti, S. Kulshreshtha, and S. Poletto,
 Floating Tunable Coupler for Scalable Quantum Computing Architectures, Phys. Rev. Applied \textbf{15}, 064063 (2021).

\bibitem{Wu}Yulin Wu, Wan-Su Bao, Sirui Cao, Fusheng Chen, Ming-Cheng Chen, Xiawei Chen, Tung-Hsun Chung,
Hui Deng, Yajie Du, Daojin Fan, Ming Gong, Cheng Guo, Chu Guo, Shaojun Guo, Lianchen Han,
Linyin Hong, He-Liang Huang, Yong-Heng Huo, Liping Li, Na Li, Shaowei Li, Yuan Li,
Futian Liang, Chun Lin, Jin Lin, Haoran Qian, Dan Qiao, Hao Rong, Hong Su, Lihua Sun,
Liangyuan Wang, Shiyu Wang, Dachao Wu, Yu Xu, Kai Yan, Weifeng Yang, Yang Yang, Yangsen Ye,
Jianghan Yin, Chong Ying, Jiale Yu, Chen Zha, Cha Zhang, Haibin Zhang, Kaili Zhang,
Yiming Zhang, Han Zhao, Youwei Zhao, Liang Zhou, Qingling Zhu, Chao-Yang Lu, Cheng-Zhi Peng,
Xiaobo Zhu, and Jian-Wei Pan, Strong Quantum Computational Advantage Using a Superconducting
 Quantum Processor, Phys. Rev. Lett. \textbf{127}, 180501 (2021).

\bibitem{Martinis}F. Arute, K. Arya, R. Babbush, D. Bacon, J. C. Bardin, R. Barends, R. Biswas,
 S. Boixo, F. G. S. L. Brandao, D. A. Buell, B. Burkett, Yu Chen, Zijun Chen, B. Chiaro, R. Collins,
  W. Courtney, A. Dunsworth, E. Farhi, B. Foxen, A. Fowler, C. Gidney, M. Giustina, R. Graff,
  K. Guerin, S. Habegger, M. P. Harrigan, M. J. Hartmann, A. Ho, M. Hoffmann, T. Huang,
  T. S. Humble, S. V. Isakov, E. Jeffrey, Zhang Jiang, D. Kafri, K. Kechedzhi, Julian Kelly,
   P. V. Klimov, S. Knysh, A. Korotkov, F. Kostritsa, D. Landhuis, M. Lindmark, E. Lucero,
   D. Lyakh, S. Mandrà, J. R. McClean, M. McEwen, A. Megrant, Xiao Mi, K. Michielsen, M. Mohseni,
    J. Mutus, O. Naaman, M. Neeley, C. Neill, M. Y. Niu, E. Ostby, A. Petukhov, J. C. Platt,
     C. Quintana, E. G. Rieffel, P. Roushan, N. C. Rubin, D. Sank, K. J. Satzinger, V. Smelyanskiy,
      K. J. Sung, M. D. Trevithick, A. Vainsencher, B. Villalonga, T. White, Z. J. Yao,
      P. Yeh, A. Zalcman, H. Neven, J. M. Martinis, Quantum supremacy using a programmable
      superconducting processor, Nature  \textbf{574},  505 (2019).





\bibitem{Friis} N. Friis, O. Marty, C. Maier, C. Hempel, M. Holzäpfel, P.
Jurcevic, M. B. Plenio, M. Huber, C. Roos, R. Blatt, and B.Lanyon,
Observation of Entangled States of a Fully Controlled 20-qubit System, Phys. Rev. X \textbf{8}, 021012 (2018).

\bibitem{Mooney}G.J. Mooney, G. A. L. White, C. D. Hill, and L. C.
L. Hollenberg, Whole-device entanglement in a 65-qubit
superconducting quantum computer, arXiv:2102.11521
(2021).


\bibitem{Zhang}Xu Zhang, Wenjie Jiang, Jinfeng Deng, Ke Wang, Jiachen Chen, Pengfei Zhang,
 Wenhui Ren, Hang Dong, Shibo Xu, Yu Gao, Feitong Jin, Xuhao Zhu, Qiujiang Guo, Hekang Li,
 Chao Song, Zhen Wang, Dong-Ling Deng, H. Wang,  Observation of a symmetry-protected topological
 time crystal with superconducting qubits, https://arxiv.org/abs/2109.05577.



\bibitem{Zhao}P. Zhao, D. Lan, P. Xu, G.M. Xue, M. Blank, X.S. Tan, H.F. Yu, and Y. Yu,
Suppression of Static ZZ Interaction in an All-Transmon Quantum Processor, Phys. Rev. Applied \textbf{16}, 024037 (2021).



\bibitem{Tan} Y. Xu, J. Chu, J. Yuan, J. Qiu, Y. Zhou, L. Zhang,
X. Tan, Y. Yu, S. Liu, J. Li, F. Yan, and D. Yu, Highfidelity,
high-scalability two-qubit gate scheme for superconducting
qubits, Phys. Rev. Lett. \textbf{125}, 240503 (2020).




\bibitem{Sung}
Y. Sung, L. Ding, J. Braumüller, A. Veps\"{a}l\"{a}inen, B. Kannan,
M. Kjaergaard, A. Greene, G. O. Samach, C. McNally, D.
Kim, A. Melville, B. M. Niedzielski, M. E. Schwartz, J. L.
Yoder, T. P. Orlando, S. Gustavsson, and W. D. Oliver,
Realization of High-Fidelity CZ and ZZ-Free ISWAP
Gates with a Tunable Coupler, Phys. Rev. X \textbf{11}, 021058
(2021).


  \bibitem{Mundada} P. Mundada, Gengyan Zhang, T. Hazard, and A. Houck,
  Suppression of qubit Crosstalk in a Tunable Coupling Superconducting Circuit,
   Phys. Rev. Applied \textbf{12}, 054023 (2019).


\bibitem{Barends} R. Barends, C. M. Quintana, A. G. Petukhov, Yu Chen, D. Kafri,
 K. Kechedzhi, R. Collins, O. Naaman, S. Boixo, F. Arute, K. Arya, D. Buell, B. Burkett,
 Z. Chen, B. Chiaro, A. Dunsworth, B. Foxen, A. Fowler, C. Gidney, M. Giustina, R. Graff,
  T. Huang, E. Jeffrey, J. Kelly, P. V. Klimov, F. Kostritsa, D. Landhuis, E. Lucero,
  M. McEwen, A. Megrant, X. Mi, J. Mutus, M. Neeley, C. Neill, E. Ostby, P. Roushan,
  D. Sank, K. J. Satzinger, A. Vainsencher, T. White, J. Yao, P. Yeh, A. Zalcman,
  H. Neven, V. N. Smelyanskiy, J. M. Martinis, Diabatic Gates for Frequency-Tunable
   Superconducting qubits, Phys. Rev. Lett. \textbf{123}, 210501 (2019).

\bibitem{Cleland} Z. L. Wang, Y. P. Zhong, L.J. He, H. Wang, J. M. Martinis, A. N. Cleland,
and Q. W. Xie, Quantum state characterization of a fast tunable superconducting resonator,
Appl. Phys. Lett. \textbf{102}, 163503 (2013).

\bibitem{Sandberg} M. Sandberg, C. M. Wilson, F. Persson, T. Bauch, G. Johansson, V. Shumeiko,
 T. Duty, and P. Delsing, Tuning the field in a microwave resonator faster than the photon lifetime,
Appl. Phys. Lett. \textbf{92}, 203501 (2008).



\bibitem{Yamaji}T. Yamaji, S. Kagami, A. Yamaguchi, T. Satoh, K. Koshino, H. Goto, Z. R. Lin,
Y. Nakamura, and T. Yamamoto, Spectroscopic observation of the crossover from a classical Duffing oscillator
to a Kerr parametric oscillator,  Phys. Rev. A \textbf{105}, 023519 (2022)


\bibitem{Wustmann1} W. Wustmann and V. Shumeiko, Parametric resonance in tunable superconducting
cavities, Phys. Rev. B \textbf{87}, 184501, (2013).


 \bibitem{Delsing} F.C. Lombardo, F.D. Mazzitelli, A. Soba, and P.I. Villar,
  Dynamical Casimir effect in a double tunable superconducting circuit, Phys. Rev. A \textbf{98}, 022512 (2018)




\bibitem{IDA}  I.-M. Svensson, Tunable superconducting resonators: Subharmonic oscillations and manipulation
of microwaves, Doctoral thesis, Chalmers University of Technology (2018).

\bibitem{Leroux} C. Leroux,  A. D. Paolo, and A. Blais, Superconducting coupler
with exponentially large on-off ratio, Phys. Rev. Applied \textbf{16}, 064062 (2021).

\bibitem{Schuster} J. Koch, T. M. Yu, J. Gambetta, A. A. Houck, D. I. Schuster,
J. Majer, A. Blais, M. H. Devoret, S. M. Girvin, and R. J. Schoelkopf,
Charge-insensitive qubit design derived from the Cooper pair box, Phys. Rev. A \textbf{76}, 042319 (2007).



 \bibitem{Sandberg1}
M. Wallquist, V. S. Shumeiko, and G. Wendin,  Selective coupling of superconducting charge
 qubits mediated by a tunable stripline cavity, Phys. Rev. B \textbf{74}, 224506 (2006).



\bibitem{Pople}  R. Krishnan and J. A. Pople, Approximate fourth-order perturbation
theory of the electron correlation energy, International Journal of Quantum Chemistry \textbf{14}, 91 (1978).

\bibitem{Zhu} Guanyu Zhu, D. G. Ferguson, V. E. Manucharyan, and J. Koch, Circuit QED
with fluxonium qubits: Theory of the dispersive regime, Phys. Rev. B \textbf{87}, 024510 (2013).


\end{thebibliography}
\end{document}